\documentclass[a4paper,11pt]{article}
\usepackage{./preamble_stuff/preamble}

\usepgfplotslibrary{external} 
\pgfrealjobname{KTequations}

\begin{document}

\title{\boldmath Khuri-Treiman equations for $3\pi$ decays of particles with spin}
\author[a,1]{M.~Albaladejo}
\emailAdd{albalade@jlab.org}

\author[b,c,2]{D.~Winney}
\emailAdd{dwinney@iu.edu}

\author[d]{I.~V.~Danilkin}

\author[e]{C.~Fern\'andez-Ram\'irez}

\author[f]{V.~Mathieu}

\author[g]{M.~Mikhasenko}

\author[h,i]{A.~Pilloni}

\author[e]{J.~A.~Silva-Castro}

\author[a,b,c]{A.~P.~Szczepaniak}


\affiliation[a]{\jlab}
\affiliation[b]{\ceem}
\affiliation[c]{\indiana}
\affiliation[d]{\mainz}
\affiliation[e]{\icn}
\affiliation[f]{\ucm}
\affiliation[g]{\cern}
\affiliation[h]{\ect}
\affiliation[i]{\genova}

\collaboration{JPAC Collaboration}
\collaborationImg{\includegraphics[height=2cm,keepaspectratio]{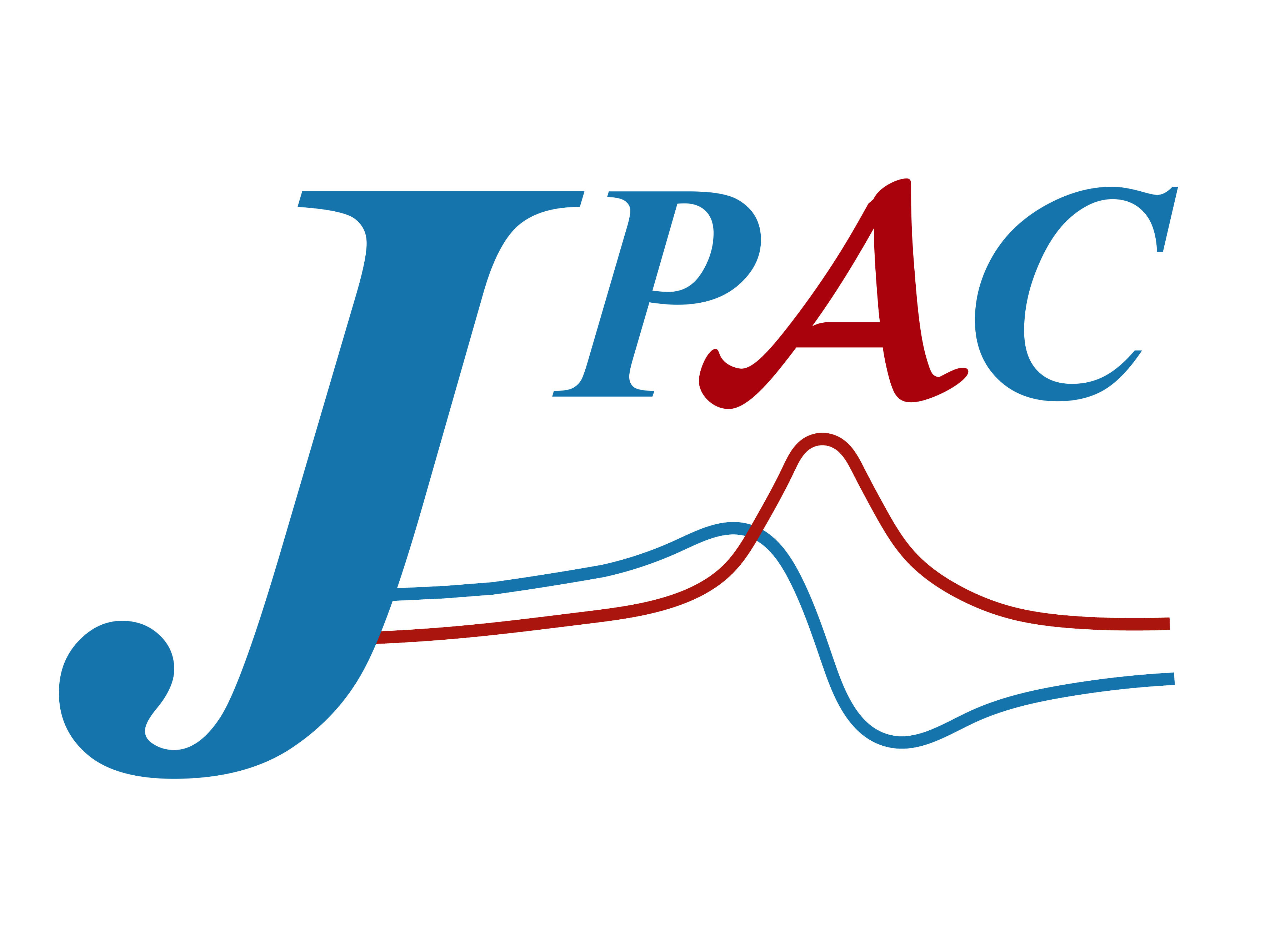}}
\preprint{JLAB-THY-19-3066}

\newcommand{\ceem}{Center for  Exploration  of  Energy  and  Matter,
Indiana  University,
Bloomington,  IN  47403,  USA}
\newcommand{\indiana}{Physics  Department,
Indiana  University,
Bloomington,  IN  47405,  USA}
\newcommand{\jlab}{Theory Center,
Thomas  Jefferson  National  Accelerator  Facility,
Newport  News,  VA  23606,  USA}
\newcommand{\icn}{Instituto de Ciencias Nucleares, 
Universidad Nacional Aut\'onoma de M\'exico, Ciudad de M\'exico 04510, Mexico}
\newcommand{\hiskp}{Universit\"at Bonn,
Helmholtz-Institut f\"ur Strahlen- und Kernphysik, 53115 Bonn, Germany}
\newcommand{\ect}{European Centre for Theoretical Studies in Nuclear Physics and related Areas (ECT$^*$) and Fondazione Bruno Kessler, Villazzano (Trento), I-38123, Italy}
\newcommand{\genova}{INFN Sezione di Genova, Genova, I-16146, Italy}
\newcommand{\cern}{CERN, 1211 Geneva 23, Switzerland}
\newcommand{\ucm}{Departamento de F\'isica Te\'orica, Universidad Complutense de Madrid, 28040 Madrid, Spain}
\newcommand{\mainz}{Institut f\"ur Kernphysik \& PRISMA$^+$  Cluster of Excellence, Johannes Gutenberg Universit\"at,  D-55099 Mainz, Germany}
\abstract{%
Khuri-Treiman equations have proven to be a useful theoretical tool in the analysis of 3-body decays, specially into the $3\pi$ final state. In this work we present in full detail the necessary generalization of the formalism to study the decays of particles with arbitrary spin, parity, and charge conjugation. To this extent, we find it most convenient to work with helicity amplitudes instead of the so-called invariant amplitudes, specially when dealing with the unitarity relations. The isobar expansions in the three possible ($s$-, $t$-, and $u$-) final channels are related with the appropriate crossing matrices. We pay special attention to the kinematical singularities and constraints of the helicity amplitudes, showing that these can be derived by means of the crossing matrix.
}

\frenchspacing
\maketitle
\newpage

\section{Introduction}\label{sec:introduction}

As our knowledge of the hadron spectrum increases, so does the complexity of the necessary  experiments and the need for better theoretical methods to analyze the experimental  data. Three-body spectra from 2-to-3 scattering or 1-to-3 decays are a fundamental tool to study both electroweak and strong physics. For example, the XYZ phenomena observed in the spectrum of charmonia are almost exclusively seen in strong decays to three particle final states~\cite{Tanabashi:2018oca}. Furthermore, constraining strong amplitudes describing three hadron final states {\it e.g.} in heavy flavor meson  or $\tau$ lepton decays is crucial in testing various BSM scenarios references~\cite{Tanabashi:2018oca}. Finally reaction amplitudes describing three-hadron interactions will soon be needed for analysis of lattice QCD simulations which are gearing towards simulations of three particle  spectra \cite{Hansen:2015zga,Briceno:2017tce,Jackura:2019bmu}.

The presence of hadrons in any reaction complicates the analysis of the amplitude of the process because of the final state interactions (FSI) of the former. In a three-body hadronic decay $X \to A\, B\, C$ there are three possible two-hadron subsystems, $AB$, $BC$, $CA$. If, for any reason, one can ignore the FSI (and the resonant content) of two of those channels, one can perform a resumation of the interaction of the remaining channel, as it is often done in 2-to-2 scattering processes. This is not feasible in many three-body decays, and certainly not possible in $3\pi$ decays, which is an important decay channel for many resonances.

Among three-body hadron decays, one of the first reactions to be studied was the $K \to 3\pi$ decay. In Ref.~\cite{Khuri:1960zz} Khuri and Treiman (KT) proposed for this reaction a dispersive approach with a set of approximations, leading to a system of linear integral equations for functions depending on a single Mandelstam variable and which build up the full amplitude. Soon after, several papers appeared discussing different aspects of the formalism \cite{Bronzan:1963mby,Aitchison:1965zz,Aitchison:1966lpz,Pasquier:1968zz,Pasquier:1969dt} and, in particular, the first application to the decay $\eta \to 3\pi$ \cite{Neveu:1970tn} (see also the recent lectures in Ref.~\cite{Aitchison:2015jxa}). Since its appeareance, the KT formalism has been applied to various three-body decay channels of both light and heavy mesons~\cite{Kambor:1995yc,Anisovich:1996tx,Descotes-Genon:2014tla,Guo:2015zqa,Guo:2016wsi,Colangelo:2016jmc,Niecknig:2012sj,Danilkin:2014cra,Isken:2017dkw,Niecknig:2015ija,Pilloni:2016obd,Albaladejo:2017hhj,Niecknig:2017ylb,Colangelo:2018jxw,Gasser:2018qtg}. The extension of the formalism to include coupled channels was achieved in Ref.~\cite{Albaladejo:2017hhj} (see also Ref.~\cite{Oller:2019opk}). Besides the practical applications, some theoretical studies about the formalism itself have also been presented. For the decay of light mesons, for instance, KT equations can be justified in chiral perturbation theory ($\chi$PT) at lowest orders via the so-called reconstruction theorem~\cite{Stern:1993rg,Zdrahal:2008bd,Bijnens:2007pr}. In Ref.~\cite{Albaladejo:2018gif} the KT formalism is applied to $\pi\pi$ scattering, finding it to be equivalent to Roy equations \cite{Roy:1971tc,Ananthanarayan:2000ht} when both formalisms are restricted to $S$- and $P$-waves, whereas when higher waves are included, good agreement is found with other dispersive approaches \cite{GarciaMartin:2011cn}.\footnote{We also point out that the KT decomposition of $\pi\pi$ scattering in Ref.~\cite{Albaladejo:2018gif} is compatible with the form of the amplitude obtained in Ref.~\cite{Yamagishi:1995kr} imposing crossing and chiral symmetries.}

An amplitude $\mA(s,t,u)$ for a scattering (2-to-2) or decay (1-to-3) process can be expanded in an infinite sum of partial wave amplitudes in one of the Mandelstam variables, say $s$. As it is well known, these partial waves are functions of a single variable, and have both right- and left-hand cut (RHC and LHC, respectively) discontinuities. Furthermore, they can contain singularities of dynamical origin, {\it i.e.}, resonances or bound states of the $s$-channel can show up as poles in the partial wave amplitudes. Because these singularities are only in the $s$-channel, those in the $t$- and $u$-channels can appear in the full amplitude $\mA(s,t,u)$ if and only if an infinite number of partial waves is kept in the partial wave expansion. Sure, this can only be done at a theoretical level, and on practical terms a truncation to a finite number of partial waves is always necessary. In this way, crossing symmetry is lost. In the KT formalism, the infinite sum of partial waves in the $s$-channel is substituted by three finite sums of so-called isobar amplitudes, one for each of the $s$-, $t$-, and $u$-channels. The isobar amplitudes are also functions of a single Mandelstam variable, but, unlike the partial wave amplitudes, they only have RHC discontinuities. In this way, the full amplitude $\mA(s,t,u)$ can have singularities coming from all the three channels, and crossing symmetry can be restored.

A quick look shows that most of the reactions to which the KT formalism has been applied are decays of pseudoscalars ($P$) or, at most, vector ($V$) mesons. As will be seen below, the KT decomposition of the decay amplitude for $V \to 3P$ is very similar to that of the decay $P \to 3P$. Therefore, it feels necessary to present a generalization of the KT formalism to the decay of mesons with any spin, parity, and charge conjugation. In this work we present such a generalization, focusing, for its ubiquity, into the decays to three pions. It will be convenient to work with helicity amplitudes~\cite{Jacob:1959at}. As mentioned before, the KT formalism allows one to keep crossing symmetry in a given amplitude. When spin appears, the crossing relations become more complicated and must be carefuly dealt with \cite{Trueman:1964zzb,MS}. We combine the standard crossing relations for helicity amplitudes with the underlying ideas of the KT formalism to obtain amplitudes KT equations for the decays of particles with spin. The crossing matrix relating the three isobar expansions serves also to derive reduced isobar amplitudes, free of kinematical singularities. These reduced isobars are thus suitable for a dispersive approach, based on the unitarity relations that we derive for them.

The structure of the manuscript is as follows. We start in Sec.~\ref{sec:kinematics} by discussing our kinematic conventions, paying special attention to the relation between the $s$- and $t$-channels center-of-mass frames. In Sec.~\ref{sec:crossing} we introduce the helicity amplitudes and its properties, in particular, crossing symmetry and how it is expressed in terms of the so-called crossing matrix. In Sec.~\ref{sec:kinsincon} we discuss the kinematical singularities and constraints of the helicity amplitudes. In Sec.~\ref{sec:KT} we obtain the KT decomposition and integral equations for the helicity amplitudes.

\section{Kinematics}\label{sec:kinematics}
\begin{figure}[t!]
\centering
\tikzsetnextfilename{Fig-schannel}%
\begin{tikzpicture}

\pgfmathsetmacro{\RX}{ 9.0};
\pgfmathsetmacro{\llx}{2.5};
\pgfmathsetmacro{\llt}{1.0};
\pgfmathsetmacro{\ths}{45};
\pgfmathsetmacro{\tht}{-60};

\node[collision] (sV) at (0,0) {};

\node (sA) at ($(sV) + (0,-\llx)$ ) {};
\node (sC) at ($(sV) + (0,+\llx)$ ) {};
\node (sB) at ($(sV) + ({\llx*sin(\ths)},{\llx*cos(\ths)})$) {};
\node (sD) at ($(sV) - ({\llx*sin(\ths)},{\llx*cos(\ths)})$) {};

\draw[pion,->-] (sA) -- (sV) node[near start,right]     {$\vec{p}_X    = + p(s) \hat{z}$};
\draw[pion,->-] (sC) -- (sV) node[near start,left]      {$\vec{p}_3  = - p(s) \hat{z}$};
\draw[pion,->-] (sV) -- (sB) node[near end,below right] {$\vec{p}_1  = + q(s) \hat{q}_s$};
\draw[pion,->-] (sV) -- (sD) node[near end,above left]  {$\vec{p}_2  = - q(s) \hat{q}_s$};

\draw[->-, very thick] ($(sV)+(0,\llt)$) arc (90:90-\ths:\llt) node[midway,above] {\; $\theta_s$};

\pgfmathsetmacro{\rxz}{2};
\pgfmathsetmacro{\zl}{0.5};

\coordinate (XZ) at ($ (sV) + (2*\rxz,-\rxz) $);
\node[draw=Black,thick,circle,minimum size=6.0pt,inner sep=0pt] (XZa) at (XZ) {};
\node[draw=Black,thick,cross=5.0pt,fill=Black, minimum size=5.0pt,inner sep=0pt] (XZb) at (XZ) {};
\draw[->,thick] (XZ) -- ($ (XZ) + (0,\zl) $) node[above] {$\hat{z}$};
\draw[->,thick] (XZ) -- ($ (XZ) + (\zl,0) $) node[right] {$\hat{x}$};

\node[collision] (tV) at ($ (sV) + (4*\rxz,0) $) {};
\node (tA) at ($(tV) + (0,-\llx)$ ) {};
\node (tC) at ($(tV) + (0,+\llx)$ ) {};
\node (tB) at ($(tV) + ({\llx*sin(\tht)},{\llx*cos(\tht)})$) {};
\node (tD) at ($(tV) - ({\llx*sin(\tht)},{\llx*cos(\tht)})$) {};

\draw[pion,->-] (tA) -- (tV) node[near start,right]     {$ \vec{p}_X^{\,\prime}   = + p(t) \hat{z}$};
\draw[pion,->-] (tC) -- (tV) node[near start,left]      {$-\vec{p}_1^{\,\prime} = - p(t) \hat{z}$};
\draw[pion,->-] (tV) -- (tB) node[near end,below left]  {$-\vec{p}_3^{\,\prime} = + q(t) \hat{q}_t$};
\draw[pion,->-] (tV) -- (tD) node[near end,above right] {$ \vec{p}_2^{\,\prime} = - q(t) \hat{q}_t$};

\draw[->-, very thick] ($(tV)+(0,\llt)$) arc (90:90-\tht:\llt) node[midway,above left] {$\theta_t$};
\end{tikzpicture}
\caption{Representation of the $s$- and $t$-channel kinematics.\label{fig:stchannel}}
\end{figure}
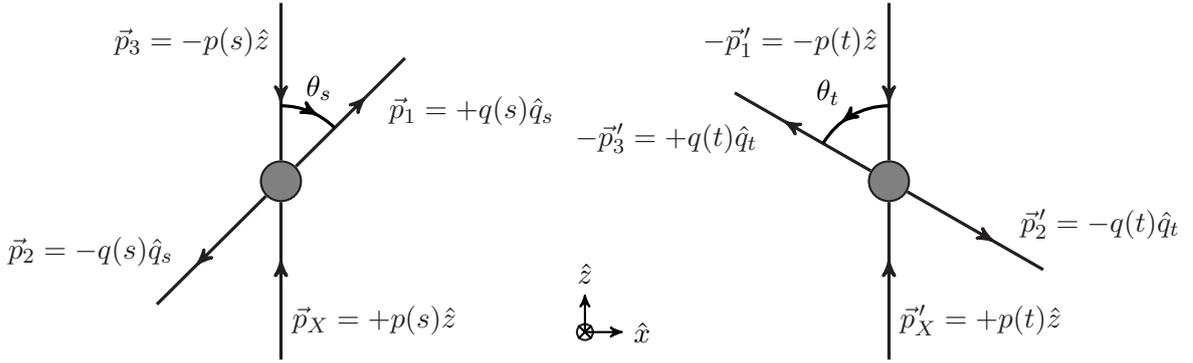
We start by discussing our conventions and notation for the kinematics of the process \footnote{Our notation, and the forthcoming discussions, are similar to those in Ref.~\cite{Hara:1971kj}.}
\begin{equation}
    \label{eq:scatter_process}
    X_J(p_X) \; \pi(p_3) \to \pi(p_1) \; \pi(p_2)
\end{equation}
where the meson $X_J$ has arbitrary quantum numbers $I$ and $J^{PC}$ (the $G$-parity quantum number, $G = -1$, is fixed by the three pion final state). Although we are primarily interested in the physical decay, $X_J\to 3\pi$, amplitudes are more easily formulated in the scattering picture, Eq.~\eqref{eq:scatter_process}, and related by crossing to the decay process through analytic continuation of momenta. For completeness, our derivations will be done at length with particular care in distinguishing kinematic quantities in different frames. This connection between different reference frames is vital in formulating a KT decomposition with the correct crossing symmetry structure.

The $X_J$ and $\pi$ masses are denoted by $M$ and $m$, respectively and invariant Mandelstam variables are defined through:
\begin{subequations}
    \label{eq:mandelstam_definitions}
    \begin{align}
    s = (p_X+ p_3)^2 &= (p_1+p_2)^2~, \nonumber\\
    t = (p_X - p_1)^2 &= (p_2-p_3)^2~,\\
    u = (p_X - p_2)^2 &= (p_3-p_1)^2~, \nonumber
    \end{align}
with
    \begin{equation}
    s+t+u = M^2 + 3 \, m^2~.
    \end{equation}
\end{subequations}
In the $s$-channel center-of-mass (CM) frame, the $z$-axis is chosen such that the four-momenta of the incoming particles have the components (see Fig.~\ref{fig:stchannel}):
\begin{subequations}
    \label{eq:kin_s}
    \begin{align}
    \label{eq:incoming_4vector}
    p_X = \left( E_{X}(s), \, +p(s) \, \hat{z} \right)
    \mand
    p_3 = \left( E_{\pi}(s), \, -p(s) \,\hat{z} \right) ~.
    \end{align}
Similarly the outgoing pions have:
    \begin{align}
    \label{eq:outgoing_4vector}
    p_1  = \left( \frac{\sqrt{s}}{2}, \; q(s) \, \hat{q}_s \right)
    \mand
    p_2  = \left( \frac{\sqrt{s}}{2}, \; -q(s) \, \hat{q}_s \right)~,
    \end{align}
\end{subequations}
where the unit vector, $\hat{q}_s$, is defined as:
    \begin{equation}
    \label{eq:vector_q_s}
    \hat{q}_s = \left( \sin\theta_s, \; 0, \; \cos\theta_s \right) ~.
    \end{equation}
The modulus of the three-momentum of the $X_J \, \pi$ and the $\pi\pi$ systems are respectively given by:
\begin{align} \label{eq:momentum_definition}
    p(s) = \sqrt{ \frac{\lao{s}}{4 \; s} }~,
    \mand
    q(s) = \sqrt{\frac{\lap{s}}{4 \;s}}~,
\end{align}
where we introduce the shorthands:
\begin{align}
    \lao{s} = \lambda(s,M^2,m^2) \mand
    \lap{s} = \lambda(s,m^2,m^2)~.
\end{align}
Above, $\lambda(x,y,z) = x^2 + y^2 + z^2 - 2(x y + y z + z x)$ is the well-known K\"all\'en or triangle function~\cite{Kallen:1964lxa}. The energies $E_{X}(s)$ and $E_\pi(s)$ in Eqs.~\eqref{eq:incoming_4vector} and \eqref{eq:outgoing_4vector} are given by:
    \begin{align}
        E_{X}(s) = \frac{s + M^2 - m^2}{2 \, \sqrt{s}}
        \mand
        E_{\pi}(s) = \frac{s + m^2 - M^2}{2 \, \sqrt{s}}~.
    \end{align}
From Eqs.~\eqref{eq:incoming_4vector} and~\eqref{eq:outgoing_4vector}, we obtain the cosine of the scattering angle $\theta_s$,
    \begin{equation} \label{eq:costhetas}
    \cos\theta_s = \frac{t-u}{4 \, p(s) \, q(s)}~.
    \end{equation}

As mentioned above, we may analogously define kinematics in the $t$-channel CM frame, i.e. the $X_J(p_X^\prime) \, \pi(-p^\prime_1) \to \pi(p^\prime_2) \, \pi(-p^\prime_3)$ process. The prime over the four-momenta is introduced to distinguish the four-vectors evaluated in the $t$-channel instead of in the $s$-channel, c.f. Eqs.~\eqref{eq:incoming_4vector} and~\eqref{eq:outgoing_4vector}. In this frame, the four-momenta for the incoming particles are given by (see Fig.~\ref{fig:stchannel}):
\begin{subequations}
    \label{eq:kin_t}
    \begin{equation}
    p_X' = \left( E_{X}(t), \;+p(t) \, \hat{z} \right)~,
    \quad
    -p'_1  = \left( E_{\pi}(t), \; -p(t) \, \hat{z} \right)~,
    \end{equation}
and for the outgoing particles
    \begin{align}
    -p'_3 =\left( \frac{\sqrt{t}}{2}, \; +q(t) \, \hat{q}_t \right)~,
    \quad
    p'_2  = \left( \frac{\sqrt{t}}{2}, \; -q(t) \, \hat{q}_t \right)~.
    \end{align}
\end{subequations}
The corresponding unit vector $\hat{q}_t$ is defined as:
    \begin{equation}
    \label{eq:vector_q_t}
    \hat{q}_t = \left(-\sin\theta_t, \; 0, \; \cos\theta_t \right)~,\\
    \end{equation}
with the cosine of the scattering angle $\theta_t$ given by:
    \begin{equation}\label{eq:costhetat}
    \cos\theta_t = \frac{s-u}{4 \, p(t) \, q(t)}~.
    \end{equation}
The relative sign of the $\hat{x}$ component of the $\hat{q}_t$ vector compared to Eq.~\eqref{eq:vector_q_s} is by convention and chosen for simplicity in later developments. This choice is related to the fact that, while the signs of $\cos\theta_s$~and~$\cos\theta_t$ in Eqs.~\eqref{eq:vector_q_s}~and~\eqref{eq:vector_q_t} are determined through kinematics ({\it e.g.}, from the product $p \, \cdot \, p_2$), {\it a priori} the signs of $\sin\theta_{s,t}$ are not fixed. In other words we have a freedom to choose the orientation of the $\hat{y}$ axis, which translates into a freedom of choosing independent $\eta_{s,t} = \pm 1$ in defining:
    \begin{align}
        \label{eqs:sins_etas}
        \sin\theta_s = \eta_s \; \sqrt{1-\cos^2\theta_s}
            = \frac{\eta_s \; \sqrt{\phi(s,t,u)}}{2 \, \sqrt{s} \, p(s) \, q(s)}
    \end{align}
and
    \begin{align}
        \sin\theta_t = \eta_t \; \sqrt{1-\cos^2\theta_t}
        = \frac{\eta_t \; \sqrt{\phi(t,s,u)}}{2 \, \sqrt{t} \, p(t) \,q(t)}~,
    \end{align}
where $\phi$ is the Lorentz-invariant Kibble function~\cite{Kibble:1960zz},
\begin{equation}
    \phi \equiv \phi(s,t,u)
    = s \, t \, u - m^2 (M^2 - m^2)^2
    = \left( 2 \, \sqrt{s} \; \sin\theta_s \, p(s) \; q(s) \right)^2 ~.
\end{equation}
The boundaries of the physical regions are defined through $\phi(s,t,u) = 0$. From the 4 four-momenta we can define an additional, fifth four-vector, that we call the Kibble vector,\footnote{Because of its transformation under parity, it is actually an axial-vector, but we shall keep the name vector since no confusion can arise.}
    \begin{equation}
    K_\mu = \epsilon_{\mu\alpha\beta\gamma} \; p_{3}^\alpha \, p_{1}^\beta \, p_{2}^\gamma~.
    \end{equation}
We note that, since none of the momenta has $\hat{y}$ component, both in the $s$- and $t$-channel CM frames $K$ can only have $\hat{y}$ component. In the $s$-channel frame then,\footnote{We are taking $\epsilon_{0123} = 1$.}
    \begin{equation}
    \label{eq:Kibble_s}
    K   = \left(0,-\sqrt{s} \, p(s) \, q(s) \sin \theta_s \; \hat{y} \right)
        = \left(0,-\eta_s \sqrt{\phi(s,t,u)} \; \hat{y} \right)~,
    \end{equation}
while in the $t$-channel frame we have:
    \begin{equation}
    \label{eq:Kibble_t}
    K ' = \left(0,-\sqrt{t} \, p(t) \, q(t) \sin \theta_t \; \hat{y} \right)
        = \left(0,-\eta_t \sqrt{\phi(t,s,u)} \; \hat{y} \right)~.
    \end{equation}

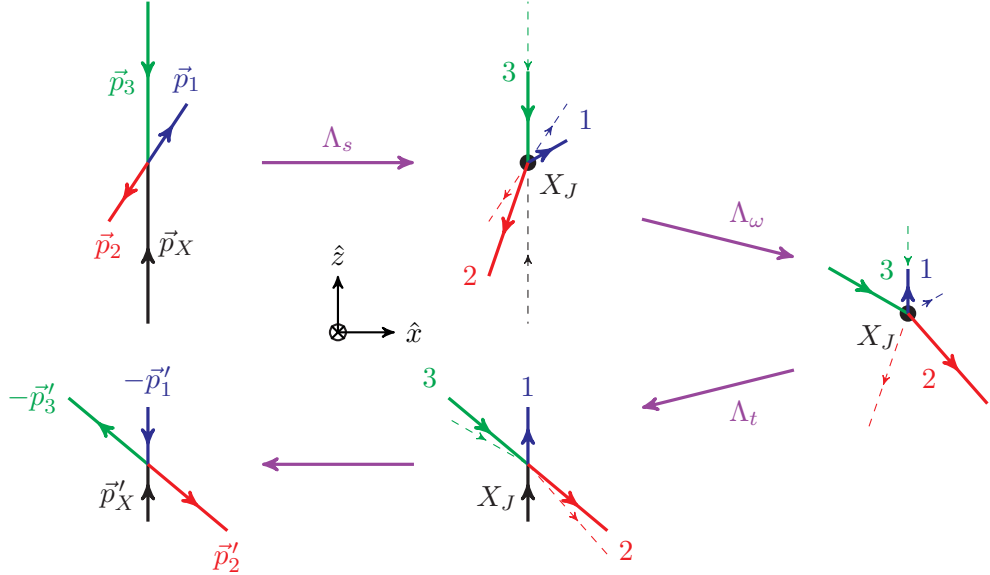
\begin{figure}[t]\centering
\tikzset{-<-/.style={decoration={
  markings,
  mark=at position .63 with {\arrow{<}}},postaction={decorate}}}
  
\tikzset{transf/.style={Purple,very thick,->}}

\tikzsetnextfilename{LoTaFig}%

\begin{tikzpicture}[scale=0.5]

\tikzset{lJ/.style={Black,very thick}}
\tikzset{l1/.style={Blue,very thick}}
\tikzset{l2/.style={Red,very thick}}
\tikzset{l3/.style={Green,very thick}}

\tikzset{llJ/.style={lJ,thin,dashed}}
\tikzset{ll1/.style={l1,thin,dashed}}
\tikzset{ll2/.style={l2,thin,dashed}}
\tikzset{ll3/.style={l3,thin,dashed}}

\coordinate (OA) at (0,0);
\coordinate (OB) at ($(OA) + (+10,+0)$);
\coordinate (OC) at ($(OB) + (+10,-4)$);
\coordinate (OD) at ($(OC) + (-10,-4)$);
\coordinate (OE) at ($(OD) + (-10,+0)$);

\coordinate (XZ) at ($ (OA) + (+5.0,-4.5) $);
\pgfmathsetmacro{\zl}{1.5};
\node[draw=Black,thick,circle,minimum size=6.0pt,inner sep=0pt] (XZa) at (XZ) {};
\node[draw=Black,thick,cross=5.0pt,fill=Black, minimum size=5.0pt,inner sep=0pt] (XZb) at (XZ) {};
\draw[->,thick] (XZ) -- ($ (XZ) + (0,\zl) $) node[above] {$\hat{z}$};
\draw[->,thick] (XZ) -- ($ (XZ) + (\zl,0) $) node[right] {$\hat{x}$};

\draw[-<-,lJ] (OA) -- ($(OA) - (+0.00,+4.27)$) node[midway,right] {$\vec{p}_X$};
\draw[-<-,l3] (OA) -- ($(OA) - (+0.00,-4.27)$) node[midway,left]  {$\vec{p}_3$};
\draw[->-,l1] (OA) -- ($(OA) + (+1.03,+1.56)$) node[above] {$\vec{p}_1$};
\draw[->-,l2] (OA) -- ($(OA) + (-1.03,-1.56)$) node[below] {$\vec{p}_2$};

\draw[-<-,lJ] (OE) -- ($(OE) - (+0.00,+1.52)$) node[above left] {$\vec{p}_X^{\,\prime}$};
\draw[->-,l3] (OE) -- ($(OE) - (+2.09,-1.76)$) node[left]  {$-\vec{p}_3^{\,\prime}$};
\draw[-<-,l1] (OE) -- ($(OE) + (+0.00,+1.52)$) node[above] {$-\vec{p}_1^{\,\prime}$};
\draw[->-,l2] (OE) -- ($(OE) + (+2.09,-1.76)$) node[below] {$\vec{p}_2^{\,\prime}$};

\begin{scope}[scale=1.0]


\draw[-<-,llJ] (OB) -- ($(OB) - (+0.00,+4.27)$) node[midway,right] {};
\draw[-<-,ll3] (OB) -- ($(OB) - (+0.00,-4.27)$) node[midway,left]  {};
\draw[->-,ll1] (OB) -- ($(OB) + (+1.03,+1.56)$) node[above] {};
\draw[->-,ll2] (OB) -- ($(OB) + (-1.03,-1.56)$) node[below] {};

\node[circle,lJ,draw,fill,inner sep=0pt,minimum size=5pt] at (OB) {};
\draw[   lJ] (OB) -- ($(OB) + (+0.00,+0.00)$) node[below right] {$X_J$};
\draw[-<-,l3] (OB) -- ($(OB) - (+0.00,-2.42)$) node[left]  {$3$};
\draw[->-,l1] (OB) -- ($(OB) + (+1.03,+0.59)$) node[above right] {$1$};
\draw[->-,l2] (OB) -- ($(OB) + (-1.03,-3.01)$) node[left] {$2$};


\draw[-<-,ll3] (OC) -- ($(OC) - (+0.00,-2.42)$) node[midway,left]  {};
\draw[->-,ll1] (OC) -- ($(OC) + (+1.03,+0.59)$) node[above] {};
\draw[->-,ll2] (OC) -- ($(OC) + (-1.03,-3.01)$) node[below] {};

\node[circle,lJ,draw,fill,inner sep=0pt,minimum size=5pt] at (OC) {};
\draw[   lJ] (OC) -- ($(OC) + (+0.00,+0.00)$) node[below left] {$X_J$};
\draw[-<-,l3] (OC) -- ($(OC) - (+2.09,-1.21)$) node[midway,above right]  {$3$};
\draw[->-,l1] (OC) -- ($(OC) + (+0.00,+1.18)$) node[right] {$1$};
\draw[->-,l2] (OC) -- ($(OC) + (+2.09,-2.39)$) node[midway,below left] {$2$};


\draw[-<-,ll3] (OD) -- ($(OD) - (+2.09,-1.21)$) node[midway,below left]  {};
\draw[->-,ll2] (OD) -- ($(OD) + (+2.09,-2.39)$) node[midway,above right] {};

\draw[-<-,lJ] (OD) -- ($(OD) - (+0.00,+1.52)$) node[above left] {$X_J$};
\draw[-<-,l3] (OD) -- ($(OD) - (+2.09,-1.76)$) node[above left]  {$3$};
\draw[->-,l1] (OD) -- ($(OD) + (+0.00,+1.52)$) node[above] {$1$};
\draw[->-,l2] (OD) -- ($(OD) + (+2.09,-1.76)$) node[below right] {$2$};

\end{scope}

\pgfmathsetmacro{\rdv}{3.0};
\pgfmathsetmacro{\rdh}{1.5};
\pgfmathsetmacro{\rdhh}{3.0};

\draw[transf] ($(OA) + (+\rdhh,+0.0)$) -- ($(OB) - (+\rdhh,+0.0)$) node[midway, above] {$\Lambda_s$};
\draw[transf] ($(OB) + (+\rdv,-\rdh)$) -- ($(OC) - (+\rdv,-\rdh)$) node[midway, above right] {$\Lambda_\omega$};
\draw[transf] ($(OC) + (-\rdv,-\rdh)$) -- ($(OD) - (-\rdv,-\rdh)$) node[midway, below right] {$\Lambda_t$};
\draw[transf] ($(OD) + (-\rdhh,+0.0)$) -- ($(OE) - (-\rdhh,+0.0)$) node[midway, below right] {};

\end{tikzpicture}
\caption{Schematic representation of the succesive Lorentz transformations in Eq.~\eqref{eq:lambdadecompos} (for the case $\eta_s = \eta_t=1$). The first and last subfigures respectively correspond to the $s$- and $t$-channel CM frames given by Eqs.~\eqref{eq:kin_s}~and~\eqref{eq:kin_t}, and shown in the left and right subfigures of Fig.~\ref{fig:stchannel}. In each of the intermediate figures the dashed lines correspond to the three-momenta prior to the applied Lorentz transformation. The last step is not a transformation but a reversal of some of the shown momenta, to match Eq.~\eqref{eq:kin_t} and Fig.~\ref{fig:stchannel} (right).\label{Fig:LoTa}}
\end{figure}

Covariant quantities in the $s$- and $t$-channel CM frames may be related to each other by a Lorentz transformation. In other words there exists a $\Lambda$ satisfying:
\begin{equation}
    \left\{ p'_1 , p'_2 , p'_3 , p_X', K'\right\}_\mu = {\Lambda_{\mu}}^{\nu} \left\{ p_1 , p_2 , p_3 , p_X, K\right\}_\nu~,
\end{equation}
where the four-vectors on the l.h.s. are expressed in the $s$-channel, Eqs.~\eqref{eq:kin_s}~and~\eqref{eq:Kibble_s}, and the four-vectors on the r.h.s. are expressed in the $t$-channel, Eqs.~\eqref{eq:kin_t}~and~\eqref{eq:Kibble_t}. The transformation, $\Lambda$, can be expressed as a series of simpler Lorentz transformations written compactly as:
\begin{equation}\label{eq:lambdadecompos}
    \Lambda = \Lambda_t \, \Lambda_{\eta_t} \, \Lambda_\omega \, \Lambda_{\eta_s} \, \Lambda_s~.
\end{equation}
The transformations $\Lambda_s$ and $\Lambda_t$, defined as:
\begin{align}
    \Lambda_s & =
    \begin{pmatrix}
    \gamma(s) & 0 & 0 & -\gamma(s) \beta(s) \\
    0        & 1 & 0 & 0 \\
    0        & 0 & 1 & 0 \\
    -\gamma(s)\beta(s) & 0 & 0 & \gamma(s)
    \end{pmatrix} ~,
    \qquad
    \Lambda_t  =
    \begin{pmatrix}
    \gamma(t) & 0 & 0 & +\gamma(t) \beta(t) \\
    0        & 1 & 0 & 0 \\
    0        & 0 & 1 & 0 \\
    +\gamma(t)\beta(t) & 0 & 0 & \gamma(t)
    \end{pmatrix} ~,
\end{align}
with
\begin{equation}
    \gamma(s) = \frac{E_X(s)}{M_X} \mand
    \beta(s) = \frac{p(s)}{E_X(s)}~,
\end{equation}
are boosts along the $\hat{z}$-axis connecting the $s$- and $t$-channel CM frames respectively  with a frame in which $X_J$ is at rest. The transformations
\begin{align}
    \label{eq:lam_eta}
    \Lambda_{\eta_s} = \diag(1, \,\eta_s, \, \eta_s, 1)~,
    \qquad
    \Lambda_{\eta_t}  = \diag(1, \, \eta_t, \, \eta_t, 1)~,
\end{align}
represent rotations of $\pi$ radians around the $\hat{z}$-axis when $\eta_s$ and/or $\eta_t$ are $-1$, and correspond to the identity transformation when $\eta_s$ and/or $\eta_t$ are $+1$. They take into account the arbitrary phases introduced in Eq.~\eqref{eqs:sins_etas}. Finally,
\begin{align}
    \Lambda_\omega & =
    \begin{pmatrix}
    1 & 0 & 0 & 0 \\
    0 & \cos\omega & 0 & -\sin\omega \\
    0 & 0 & 1 & 0 \\
    0 & \sin\omega & 0 & \cos\omega
    \end{pmatrix}
\end{align}
is a counterclockwise rotation around the $\hat{y}$-axis by an angle $\omega$ given by
\begin{align}
\label{eq:omega}
\cos \omega =
\frac{n(s,t)}{4 \, \sqrt{s t} \; p(s) \, p(t)}~, \mand
\sin \omega = \frac{M \sqrt{\phi(s,t,u)}}{2 \, \sqrt{st} \; p(s)p(t)}~,
\end{align}
with
\begin{equation}
    n(s,t) = (s + M^2 - m^2)(t + M^2 - m^2) - 2M^2(M^2-m^2)~.
\end{equation}
Fig.~\ref{Fig:LoTa} presents a schematic representation of each transformation of the total Lorentz transformation in Eq.~\eqref{eq:lambdadecompos}. As we will see in Sec.~\ref{sec:crossing}, the angle $\omega$ in Eq.~\eqref{eq:omega} will appear as the argument of additional Wigner $d$-functions in the crossing relations when considering decays with spin.

It is straightforward to check that $\det(\Lambda) = 1$, since all the matrices appearing in the product in Eq.~\eqref{eq:lambdadecompos} have unit determinant. We note the determinant of Eq.~\eqref{eq:omega} additionally gives the relation:
\begin{equation}\label{eq:nphirelation}
n^2(s,t) + 4 \; M^2 \, \phi(s,t,u) = \lambda_X(s) \, \lambda_X(t)~.
\end{equation}Computing explicitly the matrix elements of $\Lambda$, it can be seen that the only finite $(s,t)$ point where the off-diagonal ones vanish is $s=t=\frac{M^2-m^2}{2}$.\footnote{This value corresponds to $u=4m^2$. It can be seen that this point lies in the boundary of the $M \to 3m$ decay region.} At this point,
\begin{equation}
    \Lambda = \text{diag} \left( 1,\eta_s \eta_t, \eta_s \eta_t,1 \right)~.
\end{equation}
The most natural option is thus to take $\eta_s = \eta_t (=1)$, because then we can always connect the transformation between the two frames with the identity. This is indeed the convention we use in this manuscript. In this case,
\begin{equation}
\label{eq:path_endpoint}
\sqrt{s} \; p(s) \,  q(s) \; \sin\theta_s = \sqrt{t} \; p(t)\, q(t)  \; \sin\theta_t~.
\end{equation}
If we were to take $\eta_s = - \eta_t$, then $\Lambda(s=t=\frac{M^2-m^2}{2}) = \text{diag} (1,-1,-1,1)$, which corresponds to a rotation of $\pi$ around the $\hat{z}$-axis. In any case, the final Khuri-Treiman equations for the $X_J \, \pi \to \pi\pi$ amplitudes shall be independent of this convention.

\section{Helicity amplitudes}\label{sec:crossing}
We may denote the transition amplitude for the process Eq.~\eqref{eq:scatter_process} by:
    \begin{equation} \label{master_amp}
        \mA^{abcd}(\epsilon(p_X), p_3; p_1, p_2) =
        \langle \pi^c(p_1)\pi^d(p_2)
        \, | \, \hat{T} \, | \,
        X_J^a(\epsilon(p_X)) \; \pi^b(p_3) \rangle
    \end{equation}
which can be decomposed into $s$-channel helicity amplitudes,
    \begin{equation}\label{eq:hel_amp_s}
        \mA^{(s)abcd}_{\lambda}(s,t,u) = \mA^{abcd}(\epsilon^{(s)}_\lambda(p_X),p_3;p_1,p_2)~,
    \end{equation}
determined by polarization tensor $\epsilon^{(s)}_\lambda(p)$ and the helicity projection of the decaying meson, $\lambda = -J, \, \ldots, \, J$ in the $s$-channel CM frame .
For $J=1$, the components of $\epsilon^{(s)}_\lambda$ are given by Eqs.~\eqref{eq:pol_vec_J1}. Higher rank polarization tensors may also be constructed from Eq.~\eqref{eq:pol_vec_J1} but analytic forms of the polarization tensor will not be necessary for the construction of amplitudes in the helicity representation [c.f. Eq.~\eqref{eq:pol_vec_genJ}].

For physical decays the helicity amplitudes in Eq.~\eqref{eq:hel_amp_s} are projected into a basis of definite isospin in the $s$-channel through the projectors (here we assume $X$ is an isovector for concreteness; the isoscalar case is treated specifically in Sec.~\ref{sec:KT} but for the purposes of discussing symmetries it is identical to the $I=1$ amplitude):
\begin{subequations}
    \label{eq:iso_projectors}
    \begin{align}
    P^{(0)}_{abcd} &= \frac{1}{3} \delta_{ab}\delta_{cd},
    \nonumber \\
    P^{(1)}_{abcd} &= \frac{1}{2} (\delta_{ac}\delta_{bd} - \delta_{ad}\delta_{bs}),
    \\
    P^{(2)}_{abcd} &= \frac{1}{2} (\delta_{ac}\delta_{bd} + \delta_{ad}\delta_{bs}) - \frac{1}{3} \delta_{ab}\delta_{cd}
    \nonumber
    \end{align}
with
    \begin{equation}
        \label{eq:isospin_amplitudes}
        \mA_{\lambda}^{(I)}(s,t,u) \equiv \frac{1}{(2I+1)}\sum_{a,b,c,d} P^{(I)}_{abcd}
        \; \mA_{\lambda}^{(s)abcd}(s,t,u)~.
    \end{equation}
\end{subequations}
To avoid confusion with later discussion of crossing symmetry, we define and use only isospin definite amplitudes in the $s$-channel as in Eq.~\eqref{eq:isospin_amplitudes}.
Finally, the isospin amplitudes are expanded into partial waves as
    \begin{equation}
        \label{eq:isospin_pw_expansion}
        \mA_{\lambda}^{(I)}(s,t,u) = \sum_{j \geqslant |\lambda|}^\infty (2j+1) \; d^j_{\lambda 0}(\theta_s) \; A^{(I)}_{j\lambda}(s)~,
    \end{equation}
where $\theta_s \equiv \theta_s(s,t,u)$ is the $s$-channel center-of-mass scattering angle given in Eqs.~\eqref{eq:costhetas} and \eqref{eqs:sins_etas}, and $d^j_{\lambda0}(\theta_s)$ are Wigner $d$-functions.\footnote{For completeness, we specify that our convention for the latter are such that the rotation operator matrix elements are:
\begin{equation*}
    \mathcal{D}^{(j)}_{m'm}(\alpha,\beta,\gamma) = \Braket{jm'}{R(\alpha,\beta,\gamma)}{jm} = \Braket{jm'}{e^{-i\alpha \hat{J}_z} e^{-i\beta \hat{J}_y} e^{-i \gamma \hat{J}_z} }{jm} = e^{-i\alpha m'} d^{j}_{m'm}(\beta) e^{-i\gamma m}~.
\end{equation*}
For instance, $d^{1}_{01}(\beta) = \sin\beta / \sqrt{2}$.}
\subsection{Crossing symmetry}
As previously mentioned, the crossing properties between the $s$- and $t$-channel scattering (physical) regions are needed as they will be the basis for the isobar decomposition in Sec.~\ref{sec:KT}. To this end we define the $t$-channel helicity amplitudes in analogy to  Eq.~\eqref{eq:hel_amp_s} by:
    \begin{equation}
    \label{eq:hel_amp_t}
    \mA^{(t)acbd}_{\lamp}(t,s,u) = \mA^{abcd}({\epsilon^\prime}^{(t)}_\lamp(p'_X), -p^\prime_1, p^\prime_2, -p^\prime_3)~.
    \end{equation}
We remind the reader that primes denote quantities evaluated in the $t$-channel such that ${\epsilon^\prime}^{(t)}_\lamp$ is the polarization tensor for $t$-channel helicity $\lamp$, with components also given by Eqs.~\eqref{eq:pol_vec_J1}, with $\modk \to p(t)$ and $E_X \to E_X(t)$. We also note that Eq.~\eqref{eq:hel_amp_t} is a definition, and that the order of the arguments and of the isospin indices $\mA^{(t)acbd}$ is an arbitrary choice, made to simplify future formulae.

The amplitudes in Eqs.~\eqref{eq:hel_amp_s}~and~\eqref{eq:hel_amp_t} then are related by crossing symmetry \cite{Jacob:1959at,Trueman:1964zzb,Hara:1971kj,MS}, and we see that complications because of the spin arise due to the transformation of the polarization vector. We may express the $s$-channel polarization vector in the $t$-channel frame with the Lorentz transformation identified in Eq.\eqref{eq:lambdadecompos}, ${{\epsilon^\prime}_\lambda^{(s)}} = \Lambda \, \epsilon_{\lambda}^{(s)}$ as discussed in Sec.~\ref{sec:kinematics}. However this is not the same as the polarization vector initially defined in the $t$-channel, ${\epsilon^\prime}_{\lamp}^{(t)}$, because unlike the momenta four-vectors, there is additional dependence on the helicities, $\lambda$ and $\lamp$ which are defined with respect to their \textit{original} scattering kinematics and require an additional transformation. Thus with our conventions and definitions, the crossing relation reads:
\begin{subequations}    \label{eq:crossing_relation}
\begin{equation}
        \mA^{(t)acbd}_{\lamp}(t,s,u) =
        \sum_{\lambda} d^J_{\lambda\lamp}(\omega_t) \; \mA^{(s)abcd}_{\lambda}(s,t,u)~,
\end{equation}
or, equivalently,
\begin{equation}
        \mA^{(s)abcd}_{\lambda}(s,t,u) =
        \sum_{\lambda'} (-1)^{\lambda' - \lambda} \; d^J_{\lambda'\lambda}(\omega_t) \; \mA^{(t)acbd}_{\lambda'}(t,s,u)~,
\end{equation}\end{subequations}
where the angle $\omega_t \equiv \omega(s,t,u)$ has been defined in Eqs.~\eqref{eq:omega}. These relations were originally deduced by Trueman and Wick \cite{Trueman:1964zzb} (see also Ref.~\cite{MS} and, in particular, Ref.~\cite{Hara:1971kj}). They can also be directly deduced from the definitions of the amplitudes, Eq.~\eqref{eq:hel_amp_s}~and~\eqref{eq:hel_amp_t}. Finally, we also mention that these relations are obtained for some particular examples in Appendix~\ref{app:examples} in connection with the relation between helicity amplitudes and invariant functions.

The relation in Eq.~\eqref{eq:crossing_relation} is between $\mA^{(s)abcd}_{\lambda}(s,t,u)$ and $\mA^{(t)acbd}_{\lamp}(t,s,u)$. Now, the next, crucial step is to relate $\mA^{(t)acbd}_{\lambda}(t,s,u)$ with some $\mA^{(s){a'b'c'd'}}_{\lambda}(t,s,u)$ pinning down the appropriate prefactors, such that the expansion of $\mA^{(t)acbd}_{\lamp}(t,s,u)$ into partial waves does not involve new functions ${A^\prime}^{(I)}_{j\lambda}$, but the already introduced $s$-channel partial waves, $A^{(I)}_{j\lambda}$. With our conventions, the sought relation is:
    \begin{equation}
    \label{eq:crossfull}
        \mA_{\lambda'}^{(t)acbd}(t,s,u) = (-1)^{\lambda'} \; \mA_{\lambda'}^{(s)acbd}(t,s,u)~.
    \end{equation}
The proof of this relation, starting from the definitions of $\mA_\lambda^{(s)}$ and $\mA_\lamp^{(t)}$, is given in Appendix~\ref{app:proofcrossfull}. In Appendix~\ref{app:examples} the relation between helicity amplitudes and invariant functions is discussed, where this expression is explicitly cross-checked for particular $X_J \to 3\pi$ decays.  Putting together Eqs.~\eqref{eq:crossing_relation}~and~\eqref{eq:crossfull}, we have the $t$-channel helicity amplitude crossing relation:
    \begin{equation} \label{eq:crossing_relation_ss_t}
        \mA^{(s)abcd}_{\lambda}(s,t,u) = \sum_{\lambda'} (-1)^{\lambda}
        \; d^J_{\lambda'\lambda}(\omega_t) \; \mA^{(s)acbd}_{\lambda'}(t,s,u)~.
    \end{equation}
With respect to the amplitude on the left, that on the right has two indices ($b$ and $c$) and two Mandelstam variables ($s$ and $t$) swapped. Using Eq.~\eqref{eq:theta_ut}, the analogous relation for the $u$-channel can be written as:
    \begin{equation}
    \label{eq:crossing_relation_ss_u}
        \mA^{(s)abcd}_{\lambda}(s,t,u) = \sum_{\lambda'}
        (-1)^{\lambda'} \;
        d^J_{\lambda'\lambda}(\omega_u) \; \mA^{(s)adcb}_{\lambda'}(u,t,s)~,
    \end{equation}
where $\omega_u(s,t,u) \equiv \omega(s,u,t)$.
Acting on both sides of Eqs.~\eqref{eq:crossing_relation_ss_t} and~\eqref{eq:crossing_relation_ss_u} with the isospin projectors [{\it cf.} Eq.~\eqref{eq:iso_projectors}] gives us the crossing relations needed for the isobar expansion.

\subsection{Discrete symmetries}
Here we discuss additional symmetries that serve to constrain the helicity amplitudes. First, the parity relation for helicity amplitudes of definite isospin is~\cite{Jacob:1959at}:
    \begin{equation}
        \label{eq:helicity_parity}
        \mA^{(I)}_{-\lambda}(s,t,u) = \eta_\pi^3 \, \eta_X \; (-1)^{\lambda} \; \mA^{(I)}_{\lambda}(s,t,u)~,
    \end{equation}
where the naturality, $\eta_X = P \, (-1)^J$ depends on the total spin, $J$, and parity quantum number, $P$, of the decaying particle, and $\eta_\pi = -1$. This relates helicities of different parity such that only $J$ or  $J+1$ helicity amplitudes are independent for natural and unnatural decays respectively. Thus we may choose to only consider those $0 \leqslant \lambda \leqslant J$.

Next, because the sum over index $j$ in Eq.~\eqref{eq:isospin_pw_expansion} represents the total angular momentum of the initial ($X_J \, \pi$) and final ($\pi\pi$) states, it is restricted, at least, to values which satisfy Bose symmetry, $j+I=\text{even}$. From Eqs.~\eqref{eq:costhetas}~and~\eqref{eqs:sins_etas}, we see that:
    \begin{align}
    \cos\theta_s(s,u,t) =-\cos\theta_s(s,t,u) \mand
    \sin\theta_s(s,u,t) = \sin\theta_s(s,t,u)~,
\end{align}
which imply:
    \begin{equation}
        \label{eq:theta_ut}
      \theta_s(s,u,t) = \pi - \theta_s(s,t,u)~.
    \end{equation}
Therefore, we can write:
    \begin{align}        \label{eq:bose_pw}
\mA_\lambda^{(I)}(s,u,t) & =
        \sum_{j} (2j+1) \; d_{\lambda0}^j(\pi - \theta_s(s,t,u)) \; A^{(I)}_{j\lambda}(s)
        \\
        & =
        (-1)^{\lambda + I}\sum_{j}  \, (2j+1) \; d_{\lambda0}^j(\theta_s(s,t,u)) \; A^{(I)}_{j\lambda}(s)~,
        \nonumber
    \end{align}
this is,
\begin{subequations}
    \begin{equation}
        \label{eq:bose_iso}
        \mA_\lambda^{(I)}(s,u,t) = (-1)^{\lambda + I} \; \mA_{\lambda}^{(I)}(s,t,u)~,
    \end{equation}
or, in terms of the amplitudes of Eq.~\eqref{eq:hel_amp_s},
    \begin{equation}
        \mA_\lambda^{(s)abcd}(s,u,t) = (-1)^\lambda \; \mA_{\lambda}^{(s)abdc}(s,t,u)~.
    \end{equation}
\end{subequations}

\section{Kinematic singularities and constraints}\label{sec:kinsincon}
The identification of kinematic singularities is particularly important within the KT formalism as the unitarity equations involve an analytic continuation between the $s$- and $t-$channel physical regions as well as the three-body decay region. Furthermore, unitarity is easily imposed within a dispersive approach, and thus kinematical singularity-free (KSF) amplitudes are best suited for our purposes.

The construction of KSF helicity amplitudes has been the focus of many papers in the past \cite{Cohen-Tannoudji:1968lnm, Jackson:1968rfn, Franklin1966, McKerrell:1968cd, Wang:1966zza, Hara:1964zza,MS,Lutz:2011xc,Stoica:2011cy}. (See further discussions in Refs.~\cite{Mikhasenko:2017rkh,Pilloni:2018kwm}.) In particular, Refs.~\cite{Cohen-Tannoudji:1968lnm} and \cite{Hara:1964zza} show that all singularities arising from the kinematics of the reaction can be identified by the analytic structure of the crossing relation, Eq.~\eqref{eq:crossing_relation}. Here we will present results for use in the derivation of KT equations and we refer to the literature for a more complete discussion of kinematic singularities.

To motivate the systematic construction of amplitudes within the Khuri-Treiman framework for the $X_J \to 3\pi$ reaction with arbitrary quantum numbers, we demonstrate this process for two cases of interest: the $a_1 \; (J^{PC} = 1^{++})$ and $\pi_2 \; (J^{PC} = 2^{-+})$. By using the crossing relation for helicity amplitudes to identify kinematic singularities, we alleviate the need to write amplitudes in terms of general Lorentz covariant structures which can prove nontrivial and unwieldy for particles with spin.
Connections to the conventional method of identifying kinematic singularities, i.e. by matching to Lorentz invariant amplitudes, are illustrated in Appendix~\ref{app:examples}.
\subsection[Example: $a_1 \; (J^{PC} = 1^{++})$]{\boldmath Example: $a_1 \; (J^{PC} = 1^{++})$}\label{sec:singul_a1}
As mentioned above, the starting point given a set of quantum numbers, $J^{PC}$, for the decaying particle is the crossing relation, Eq.~\eqref{eq:crossing_relation}. Using the parity relation, Eq.~\eqref{eq:helicity_parity}, with $J^P=1^+$ we see the process, $a_1 \, \pi \to \pi \pi$ has only two independent helicity amplitudes for which we choose $\lambda = 0, \, +1$. This allows us to write Eq.~\eqref{eq:crossing_relation} compactly as (we remove isospin indices as the kinematic singularities do not depend on the isospin projection):
  \begin{align} \label{eq:matrix-helicity_a1}
    \begin{pmatrix}
    \sqrt{2} \, \mA^{(s)}_{+}(s,t,u) \\
     \mA^{(s)}_{0}(s,t,u)
    \end{pmatrix}
    =
    \begin{pmatrix}
       \cos \omega  &  -\sin \omega   \\
       \sin \omega  & \cos \omega
    \end{pmatrix}
    \begin{pmatrix}
    \sqrt{2} \, \mA^{(t)}_{+}(t,s,u) \\
     \mA^{(t)}_{0}(t,s,u)
    \end{pmatrix}&  \nonumber \\
    \equiv 
    \mathbb{C}(s,t)\,
    \begin{pmatrix}
    \sqrt{2} \, \mA^{(t)}_{+}(t,s,u) \\
     \mA^{(t)}_{0}(t,s,u)
    \end{pmatrix}&
  \end{align}
where the angle $\omega$ is given by Eq.~\eqref{eq:omega}.
We now define
    \begin{equation}
        \label{eq:H_hat}
        \mA^{(s)}_\lambda(s,t,u) = K_{\lambda}(s,t) \; \mAhat^{(s)}_\lambda(s,t,u)~,
    \end{equation}
such that $\mAhat_\lambda^{(s)}$ has only dynamical singularities, with the function $K_\lambda(s,t)$ factoring out the kinematic singularities in direct and cross-channel variables. With these functions, we construct the diagonal kinematic function matrix $\mathbb{K}(s,t)$ by:
\begin{subequations}
\begin{equation}
        \mathbb{K}(s,t) = \diag \big( K_+(s,t), K_0(s,t) \big) \; .
\end{equation}
Then, inserting Eq.~\eqref{eq:H_hat} into Eq.~\eqref{eq:matrix-helicity_a1}, we have:
  \begin{align}
    \label{eq:kin_matrix}
    \begin{pmatrix}
    \sqrt{2} \, \mAhat^{(s)}_{+}(s,t,u) \\
     \mAhat^{(s)}_{0}(s,t,u)
    \end{pmatrix}
    =
    \mathbb{K}^{-1}(s, t)
    \, \mathbb{C}(s,t) \,
    \mathbb{K}(t, s)
    \times
    \begin{pmatrix}
    \sqrt{2} \, \mAhat^{(t)}_{+}(t,s,u) \\
     \mAhat^{(t)}_{0}(t,s,u)
    \end{pmatrix} \,
  \end{align}
\end{subequations}
Now we want to determine the functions $K_\lambda$ such that, by construction, the vectors on the l.h.s and r.h.s in Eq.~\eqref{eq:kin_matrix} have no kinematic singularities, for which the matrix product must then also be kinematic-singularity-free. Using Eqs.~\eqref{eq:omega}, we identify:
    \begin{align}
        \label{eq:a1_k_matrix}
        \mathbb{K}(s,t) &= \diag \big( s^{-1/2} \, \sin\theta_s \; \lambda^{1/2}_\pi(s), \; \lambda_X^{-1/2}(s) \big) \nonumber \\
            &= \lambda_X^{-1/2}(s) \; \diag\big( \sqrt{\phi} / 2, \; 1 \big)~,
    \end{align}
such that the elements of the crossing matrix
\begin{subequations}
  \begin{align}
    \label{eq:matrix_con_s}
    \begin{pmatrix}
    \sqrt{2} \, \mAhat^{(s)}_{+}(s,t,u) \\
     \mAhat^{(s)}_{0}(s,t,u)
    \end{pmatrix}
    =
    \frac{1}{\lambda_X(t)}
    \begin{pmatrix}
       n(s,t)  & M \\
       -4 M \phi & n(s,t)
    \end{pmatrix}
    \begin{pmatrix}
    \sqrt{2} \, \mAhat^{(t)}_{+}(t,s,u) \\
    \mAhat^{(t)}_{0}(t,s,u)
    \end{pmatrix} \,
  \end{align}
are regular in $s$ and $t$. Additionally we may invert Eq.~\eqref{eq:matrix_con_s} and use Eq.~\eqref{eq:crossfull} [together with Eq.~\eqref{eq:nphirelation}] to write everything in terms of $s$-channel helicity amplitudes:
    \begin{align}
        \label{eq:matrix_con_t}
        \begin{pmatrix}
        \sqrt{2} \, \mAhat^{(s)}_{+}(t,s,u) \\
         \mAhat^{(s)}_{0}(t,s,u)
        \end{pmatrix}
        =
        \frac{1}{\lambda_X(s)}
        \begin{pmatrix}
           -n(s,t)  & M \\
           4 M \phi & n(s,t)
        \end{pmatrix}
        \begin{pmatrix}
        \sqrt{2} \, \mAhat^{(s)}_{+}(s,t,u) \\
        \mAhat^{(s)}_{0}(s,t,u)
        \end{pmatrix} \, .
  \end{align}
\end{subequations}
Except for the factor $1/\lambda_X(s)$, the crossing matrix between the l.h.s. and the r.h.s. has no kinematics singularities. Because $\lambda_X(s) \to 0$ as $s \to (M \pm m)^2$, the r.h.s. of Eq.~\eqref{eq:matrix_con_t} would develop a singularity at threshold and pseudo-threshold. Since the l.h.s. is expanded into helicity partial waves in the $t$ variable, this singularity could not be reabsorbed into the definition of $\mAhat_\lambda^{(s)}(t,s,u)$. Thus, in order to have the l.h.s. free of kinematic singularities, the amplitudes $\mAhat_\lambda^{(s)}(s,t,u)$  on the r.h.s. cannot be independent at these points. Specifically, in order to cancel this behaviour, they must obey the kinematic constraint:
    \begin{equation}
        \label{eq:constraint_a1}
        \bigg[
        \sqrt{2} \, n(s,t) \; \mAhat^{(s)}_{+} (s,t,u) - M\, \mAhat^{(s)}_{0}(s,t,u)
        \bigg] \stackrel[s\, \sim\, (M\pm m)^2]{}{\propto} \lambda_X(s)~.
    \end{equation}

We note that the singularities in Eq.~\eqref{eq:a1_k_matrix} are identical to those tabulated in Refs.~\cite{MS, Collins} for the $a_1$ quantum numbers. This includes powers of $\big(\sin\theta_s\big)^{|\lambda|}$ corresponding to the so-called ``half-angle factors'' which contain all singularities in the cross-variable. The $s$-dependence additionally includes necessary powers of $\sqrt{s}$ to ensure the factorization of Regge poles and correct behavior of partial waves near threshold. 

We now generalize the results obtained for the $a_1$ case before. For $X_J \, \pi \to \pi\pi$, we may write:
    \begin{align}
        \label{eq:K_factor_general}
        K_\lambda(s,t)  &= \big(s^{-1/2} \sin\theta_s\big)^{|\lambda|}  \,
        \big( \lambda^{1/2}_X(s) \big)^{|\lambda| - Y_X} \, \big( \lambda^{1/2}_\pi(s) \big)^{|\lambda|} \; \nonumber \\
        &= \sqrt{(\phi / 4)^{|\lambda|}} \; \sqrt{(\lambda_X(s) )^{-Y_X}}
    \end{align}
where the exponent
    \begin{equation}
    \label{eq:Y_exp}
        Y_X = J - \frac{1}{2} \big[1 + \eta_X\big]
    \end{equation}
depends on the spin and naturality of the decaying particle to compensate for the possible mismatch between factors of momentum at the $X_J \, \pi$ vertex in the helicity and $LS$ bases.

For a crossing matrix $\mathbb{C}(s,t)$ given by the Trueman-Wick crossing relation, Eq.~\eqref{eq:crossing_relation} and kinematic-singularity matrices $\mathbb{K}(s,t)$ whose elements are given by Eq.~\eqref{eq:K_factor_general}, the matrix
\begin{subequations}
    \begin{equation}
    \label{def:C_hat}
    \widehat{\mathbb{C}}(s,t) =  \mathbb{K}^{-1}(s, t) \,
     \mathbb{C}(s,t) \,
    \mathbb{K}(t, s)
    \end{equation}
is free of kinematic singularities and is regular in all variables (except at threshold and pseudo-threshold where additional constraints may apply) and satisfies
    \begin{equation}
        \widehat{\mathbb{C}}^{-1}(s,t) = \widehat{\mathbb{C}}(t,s) \, .
    \end{equation}
\end{subequations}
\subsection[Example: $\pi_2 \; (J^{PC} = 2^{-+})$]{\boldmath Example: $\pi_2 \; (J^{PC} = 2^{-+})$} \label{sec:singul_pi2}
The $\pi_J$ family of mesons are of particular interest as candidates for QCD structure beyond typical $q\bar{q}$ states. Although the $\pi_1$ state is more established both theoretically and experimentally as a spin-exotic hybrid candidate, we focus on the $\pi_2$ as a more illuminating example for its non-trivial helicity structure.

As before, using the parity relation of Eq.~\eqref{eq:helicity_parity}, we choose the three independent helicity amplitudes to correspond to $\lambda = 0, +1, +2$. Then the analogous expression to Eq.~\eqref{eq:matrix_con_t} for the $\pi_2$ is
\begin{subequations}
  \begin{align} \label{eq:matrix-helicity_pi2}
    \begin{pmatrix}
        \Hlam{s}{2}(t,s,u) \\
        \Hlam{s}{1}(t,s,u) \\
        \Hlam{s}{0}(t,s,u)
    \end{pmatrix}
    =
    \mathbb{C}(s,t)
    \begin{pmatrix}
        \Hlam{s}{2}(s,t,u) \\
        \Hlam{s}{1}(s,t,u) \\
        \Hlam{s}{0}(s,t,u)
    \end{pmatrix} \,
  \end{align}
where
    \begin{equation}
    \mathbb{C} = \mathbb{C}(s,t) = \mathbb{C}(t,s) =
        \begin{pmatrix}
        (1 + \cos^2\omega) /2 &  - (\sin 2\omega)/2 & \sqrt{3/8} \, \sin^2\omega \\
        - (\sin 2\omega)/2 & 1- 2\cos^2\omega  & \sqrt{3/8} \,\sin2\omega \\
        \sqrt{3/2} \, \sin^2\omega &  \sqrt{3/2} \, \sin2\omega & (3\cos^2\omega - 1) /2 \\
        \end{pmatrix} \; .
    \end{equation}
\end{subequations}
Directly from Eq.~\eqref{eq:K_factor_general}, with $\eta_{\pi_2} = -1$, we write
    \begin{equation}
        \mathbb{K}_s \equiv \mathbb{K}(s,t) = \lambda^{-1}_X(s) \; \diag \bigg(\phi /4 , \; \sqrt{\phi}/2 , \; 1 \bigg)
    \end{equation}
and verify that (with $n_{st} \equiv n(s,t)$ and $\lambda_X^s \equiv \lambda_X(s)$)
  \begin{align}
  \label{eq:matrix_KCK}
    \widehat{\mathbb{C}}(t,s) = \mathbb{K}^{-1}_t \,
    & \mathbb{C} \,
    \mathbb{K}_s =
    \frac{1}{\lambda_X^2(s)}
    \begin{pmatrix}
       (\lambda_X^s\lambda_X^{t} + n_{st}^2)/2 & - M \, n_{st} & 16 \, \phi^2 \, M^2 \\
        -4 \, M  \phi \; n_{st}  & (4M^2 \, \phi - n_{st}) & \sqrt{3/2} \, M \, n_{st} \\
        8 \sqrt{6} \, M^2 \phi^2 & 4\sqrt{6} \, M \phi \; n_{st} & \; \; (n_{st} - 2 M^2 \, \phi)
    \end{pmatrix}
\end{align}
is free of singularities except at (pseudo-)threshold. The factor of $\lambda_X^2(s)$ in front gives the kinematic constraints analogous to Eq.~\eqref{eq:constraint_a1} for the $\pi_2$ as $s\to (M\pm m)^2$.
It is worth noting that the matrices in Eqs.~\eqref{eq:matrix_con_t} and \eqref{eq:matrix_KCK} are not only free of singularities, but that are their own inverse up to the interchange of $s\leftrightarrow t$. This is a manifestation of the crossing property of Eq.~\eqref{eq:crossfull} whereby helicities amplitudes in different channels are related by the interchange of variables.
\subsection{Partial-wave expansion}
In the previous subsection we have derived the KSF helicity amplitudes $\mAhat^{(s)}_\lambda(s,t,u)$ in terms of the original ones, $\mA^{(s)}_\lambda(s,t,u)$, see {\it e.g.} Eqs.~\eqref{eq:H_hat}~and~\eqref{eq:a1_k_matrix}. In this section, building upon the previous examples, we discuss additional steps required to define KSF partial wave amplitudes.

Because the kinematic structure is independent of isospin we trivially extend these results to the $s$-channel helicity amplitudes of definite isospin in Eq.~\eqref{eq:isospin_amplitudes}, where analogous to Eq.~\eqref{eq:H_hat},
    \begin{equation}
        \label{eq:H_hat_iso}
        \mA^{(I)}_\lambda(s,t,u) = K_\lambda(s, t) \; \mAhat^{(I)}_\lambda(s,t,u)
    \end{equation}
with the kinematic function being exactly the same as in Eq.~\eqref{eq:K_factor_general}.
These amplitudes may then be expanded into helicity partial waves as in Eq.~\eqref{eq:isospin_pw_expansion}.

We recall that Eq.~\eqref{eq:K_factor_general} includes appropriate half-angle factors which can be pulled out from the Wigner-$d$ function, by defining the {\it reduced} $d$-function $\hat{d}^j_{\lambda0}(\theta_s)$ as:
    \begin{equation}
        d_{\lambda0}^j(\theta_s) \equiv
        \big(\sin\theta_s\big)^{|\lambda|}
        \; \hat{d}^j_{\lambda 0}(\theta_s) .
    \end{equation}
However, because of the denominator in its definition, Eq.~\eqref{eq:costhetas}, $\cos\theta_s$ becomes singular as $s$ approaches (pseudo-)threshold at $(M\pm m)^2$. Specifically, it develops a square-root type singularity proportional to $\sqrt{(\lao{s}\lap{s})^{j - |\lambda|}}$. Writing $\mAhat_\lambda^{(I)}(s,t,u)$ from Eq.~\eqref{eq:H_hat}, we explicitly factor this pole as:
\begin{align}
    \label{littlehat}
    \mAhat^{(I)}_\lambda(s,t,u) & =
    \sum_{j\geqslant |\lambda|}^\infty (2j+1) \sqrt{(\lao{s}\lap{s})^{j - |\lambda|}}
    \;
    \frac{d^j_{\lambda 0}(\theta_s) \;  A^{(I)}_{j\lambda}(s)}
    {
     K_\lambda(s,t) \, \sqrt{(\lao{s}\lap{s})^{j - |\lambda|}}
     }
     \\
    & = \sum_{j\geqslant |\lambda|}^\infty (2j+1)
    \sqrt{(\lao{s}\lap{s})^{j - |\lambda|}}
    \;
    \Bigg[
    \frac{d^j_{\lambda 0}(\theta_s)
    }{\sin^{|\lambda|} \theta_s}
    \Bigg]
    \Bigg[\frac{\sqrt{s^{|\lambda|}} \; A^{(I)}_{j\lambda}(s)
    }{\big( \lambda^{1/2}_X(s) \big)^{j- Y_X} \, \big( \lambda^{1/2}_\pi(s) \big)^{j}}\Bigg]~. \nonumber
\end{align}
We stress that the sum over $j$ is restricted to values $j \geqslant \left\lvert \lambda \right\rvert$, and that this restriction is implicitly carried over all formulae below involving $j$ and $\lambda$.

Since $\mAhat_{\lambda}^{(I)}(s,t,u)$ is KSF, and so is the factor outside the brackets in Eq.~\eqref{littlehat}, we deduce that the expressions in brackets must also be KSF. We thus define the KSF partial wave helicity amplitude, $\Ahat_{j\lambda}^{(I)}(s)$, as:
    \begin{equation}
        \label{eq:littleh_hat}
        A_{j\lambda}^{(I)}(s) =
            \frac{\big( \lambda^{1/2}_X(s) \big)^{j- Y_X} \, \big( \lambda^{1/2}_\pi(s) \big)^{j}
            }{\sqrt{s^{|\lambda|}}
            } \Ahat_{j\lambda}^{(I)}(s)~.
    \end{equation}
The partial wave amplitude, in fact, is expected to behave as \cite{Collins}:
\begin{equation}\label{eq:thresholdbehavior}
        A^{(I)}_{j\lambda}(s) \sim \sqrt{\lambda_X^{L_i}(s)} \, \sqrt{\lambda_\pi^{L_f}(s)}
\end{equation}
near threshold where $L_i$ and $L_f$ are the minimal orbital angular momentum allowed for the helicity/parity combination of the $X\pi$ and $\pi\pi$ states respectively. This is the so-called angular momentum barrier factor for particles with spin which depends on the spin of the partial wave, corrected by the additional power $Y_X$ in the $K_\lambda$ function in Eq.~\eqref{eq:K_factor_general}.
Here we note that there are in fact three different threshold points associated with the behavior in Eq.~\eqref{eq:thresholdbehavior}: $(M \pm m)^2$ and $4m^2$ which may give rise to kinematic square-root type singularities. The orbital angular momentum of the final $\pi\pi$ system is clearly $L_f = j$, which agrees with Eqs.~\eqref{eq:thresholdbehavior} and \eqref{eq:littleh_hat}. The orbital angular momentum of the initial state with total angular momentum $j$ however obeys
    \begin{equation}
        \label{eq:L_i}
        L_i \geqslant |j- Y_X|~,
    \end{equation}
where the minimally allowed $L_i$ for $j < J - 1 \; (J)$ for natural (unnatural) decays respectively is different than those with $j \geqslant J -1 \; (J)$. Examining Eq.~\eqref{eq:littleh_hat} we see we need to account for this dependence.

We combine the half-angle factors and the behaviors in Eqs.~\eqref{eq:littleh_hat} and ~\eqref{eq:L_i} to define the kinematic function for KSF partial waves:
\begin{subequations}
    \begin{align}
            K_{j\lambda}(s, \theta_s) &=
            \big(s^{-1/2} \, \sin\theta_s \big)^{|\lambda|} \;
            \big( \lambda^{1/2}_X(s) \big)^{|j- Y_X|} \, \big( \lambda^{1/2}_\pi(s) \big)^{j}
            \nonumber \\
            &= \big(\sqrt{\phi}/2\big)^{|\lambda|}
            \; \sqrt{(\lao{s}\lap{s})^{j - |\lambda|}} \sqrt{\lao{s}^{Y_j}}
            \;
    \end{align}
with the exponent $Y_j$ now having the additional $j$ dependence for lowest partial wave  [{\it cf.} Eq.~\eqref{eq:Y_exp}]:
    \begin{equation}
        Y_j = |j-Y_X| - j~.
    \end{equation}
\end{subequations}

In general then, we may write the fully factorized amplitude as a sum of $s$-channel partial waves compactly as
    \begin{equation}
        \mA^{(I)}_\lambda(s,t,u) = \sum^{\infty}_{j\geqslant |\lambda|} (2j+1) \;
        K_{j\lambda}(s, \theta_s) \; \hat{d}_{\lambda0}^j(\theta_s)
        \;   \Ahat^{(I)}_{j\lambda}(s)
    \end{equation}
where the we see KSF partial-wave amplitudes given by
    \begin{equation}
    \label{eq:def_ksf_partialwave}
        \Ahat^{(I)}_{j\lambda}(s) = \frac{1}{2} \int d\cos\theta^\prime \;
        \big(\sin^2\theta^\prime\big)^{|\lambda|} \; \hat{d}^j_{\lambda0}(\theta^\prime) \; \frac{\mA^{(I)}_\lambda(s,t(s, \theta^\prime),u(s,\theta^\prime))}{K_{j\lambda}(s, \theta^\prime)} \,     \end{equation}
are left as dynamical functions to be constrained by the KT equations in the following section. We note here that angular momentum conservation restricts $j \geqslant \left\lvert \lambda \right\rvert$ and we thus do not need to cases with $j < \left\lvert \lambda \right\rvert$.

To end our discussion about the partial wave expansion, we note that the kinematical constraint in Eq.~\eqref{eq:constraint_a1}
is expressed in terms of the KSF full amplitude, $\mAhat$, and, upon projection of the l.h.s. into helicity partial waves [Eq.~\eqref{eq:isospin_pw_expansion}], can be expressed as kinematical constraints for the latter. When this is done, it is easy to check that Eq.~\eqref{eq:constraint_a1} is completely equivalent, up to some redefinitions, to the kinematical constraints obtained in Ref.~\cite{Mikhasenko:2017rkh}. Furthermore, the fulfillment of the kinematical constraints by the partial waves can modify the definition of the KSF isobar in Eq.~\eqref{eq:littleh_hat}, $\mAhat_{j\lambda}^{(I)}$, for lower (in $j$) partial waves, and these cases must be specifically checked.

\section{Khuri-Treiman equations}\label{sec:KT}
At the heart of the KT formalism is the isobar decomposition which truncates the infinite series of $s$-channel helicity partial wave amplitudes in favor of a finite series of isobar amplitudes in each channel. Later, elastic ($\pi\pi \to \pi \pi$) unitarity is imposed on each channel simultaneously. To treat all possible decays, $X_J \to 3\pi$, tabulated in Table~\ref{tab:allowed}, we consider two cases based on their isospin structure. Specific implementation between different mesons of the same isospin amounts only to different parity constraints as seen in Eq.~\eqref{eq:helicity_parity}.

\begin{table}[]
    \centering
    \begin{tabular}{|c|c|c|c|} \hline
 $PC$ & $J_\text{min}$ & $I$ & notation (for $I=0,1$) \\ \hline\hline
 $++$ & $1$            & odd & $a_J$ \\ \hline
 $+-$ & $1$            & even & $h_J$ \\ \hline
 $-+$ & $0$            & odd  & $\pi_J$ \\ \hline
 $--$ & $0$            & even & $\omega_J/\phi_J$ \\ \hline
    \end{tabular}
    \caption{Allowed quantum numbers of $X_J$ in $X_J \to 3\pi$ decays.\label{tab:allowed}}
\end{table}

\subsection[Isoscalar decay ($\omega_J / \phi_J$ and $h_J$)]{\boldmath Isoscalar decay ($\omega_J / \phi_J$ and $h_J$)}
\paragraph{Isobar Decomposition.}
We start with the case where the decaying particle $X_J$ has total isospin, $I = 0$ and therefore does not have additional complications from different isospin projections as the only allowed intermediate states must be isospin-1. In other words, for the $s$-channel scattering process $X_J(p) \pi(p_3) \to \pi(p_1)\pi(p_2)$, the only allowed amplitudes are
    \begin{equation}\label{eq:isoscalaramplitude}
        \mA^{(s)}_\lambda(s,t,u) \equiv \sum_{a, \, b, \, c} P_{abc} \; \mA_\lambda^{(s)abc}(s,t,u)
    \end{equation}
where $P_{abc} = -i \, \epsilon_{abc}/\sqrt{2}$ is the anti-symmetric isospin factor corresponding to the coupling of three isospin-1 pions to the isoscalar initial state. In Sec.~\ref{sec:crossing} we discussed the crossing symmetry relations for the case of an isovector decay.
For the case we are discussing here, the formulae equivalent to Eqs.~\eqref{eq:crossing_relation}~and~\eqref{eq:crossfull} are:
\begin{subequations}\label{eq:crossing_pre_isoscalar}
\begin{align}
\mA_\lambda^{(s)abc}(s,t,u) & = \sum_{\lambda'} (-1)^{\lambda' - \lambda} \,  d^J_{\lambda'\lambda}(\omega) \, \mA_{\lambda'}^{(t)bac}(t,s,u)~,
\end{align}
and
\begin{align}
\mA_\lambda^{(t)bac}(t,s,u) & = (-1)^\lambda \, \mA_\lambda^{(s)bac}(t,s,u)~.
\end{align}
\end{subequations}
Inserting these into Eq.~\eqref{eq:isoscalaramplitude}, we get:
\begin{equation}\label{eq:crossfull_isoscalar}
\mA_{\lambda}(s,t,u) = (-1)^{\lambda + 1}  \sum_{\lambda'} d^J_{\lambda'\lambda}(\omega) \, \mA_{\lambda'}(t,s,u)~.
\end{equation}
The full amplitude, $\mA_{\lambda}(s,t,u)$ is projected into partial waves as in Eq.~\eqref{eq:isospin_pw_expansion} but with the isoscalar decay condition we may drop the superscript isospin index:
\begin{equation}\label{eq:pw_isoscalar}
        \mA_{\lambda}(s,t,u) = \sum_{j \geqslant |\lambda|} (2j+1) \; d_{\lambda0}^j(\theta_s) \; A_{j\lambda}(s)~.
\end{equation}
Now we need to perform the KT decomposition of this full amplitude. As said, this consist in substituting the infinite sum of helicity partial waves above by three truncated sum of helicity isobars. The expansion reads:
    \begin{align}\label{eq:KT_isoscalar}
         \mA_\lambda (s,t,u)
         = \bar{\mA}_\lambda(s,t,u)  &+ \sum_{\lamp}  (-1)^{\lambda +1} \; d^J_{\lamp\lambda}(\omega_t) \; \bar{\mA}_\lamp(t,s,u) \\
         &+ \sum_\lamp (-1)^{\lamp +1} \; d^J_{\lamp\lambda}(\omega_u) \; \bar{\mA}_\lamp(u,t,s)~, \nonumber
    \end{align}
where, analogous to Eq.~\eqref{eq:pw_isoscalar}, each of the ``amplitudes'' $\bar{\mA}_\lambda(s,t,u)$ is defined as a truncated sum of isobars $a_{j\lambda}(s)$:
    \begin{equation}
        \label{eq:isobar_def}
        \bar{\mA}_{\lambda}(s,t,u) = \sum^\jmax_{j \geqslant |\lambda|} (2j+1) \; d_{\lambda0}^j(\theta_s) \; a_{j\lambda}(s)~.
    \end{equation}
In the following, we make some points about this expansion.

 First, by virtue of Eq.~\eqref{eq:isobar_def}, the isobar sums in Eq.~\eqref{eq:KT_isoscalar} must be truncated at the same $\jmax$ for each channel in order to maintain crossing symmetry. By adding the $t$- and $u$-channel isobar expansions, we recover some of the singularity structure in these variables corresponding to exchange physics that is lost when the infinite sum in Eq.~\eqref{eq:isospin_pw_expansion} is truncated.

 Second, the cross-channel isobar terms must include the additional rotation by $\omega_t$ and $\omega_u$ as per Eq.~\eqref{eq:crossing_relation} for particles with spin. From the discussion in Sec.~\ref{sec:crossing}, it is intuitive why Eq.~\eqref{eq:KT_isoscalar} restores crossing symmetry. For completeness, though, we present a detailed proof in Appendix~\ref{app:proof2}. We stress here that any KT-like expansion which does not take into account this structure does not preserve crossing symmetry, contrary to the original program.

Third, we make the necessary distinction that the isobar amplitudes are not the same as the partial wave amplitudes of Eq.~\eqref{eq:isospin_pw_expansion}. By assumption, each isobar amplitude only has a RHC opening from the two pion threshold as opposed to both LHC and RHC structure in the whole partial-wave. This is equivalent to assuming that the (reduced) isobar amplitude only satisfies the single-variable dispersion relation (shown here as unsubtracted but in general may have additional polynomial dependence):
    \begin{equation}
    \label{eq:dispersion_relation}
    \hat{a}_{j\lambda}(s) = \frac{1}{\pi} \int_{4m^2}^\infty ds^\prime \; \frac{\Disc \hat{a}_{j\lambda}(s^\prime)}{s^\prime - s - i\epsilon}
    \end{equation}
as opposed to the more general dispersion relation for the partial wave amplitude, that involves also an integral along the LHC.

To make this difference more concrete, we write out the full expansion in isobar amplitudes:
    \begin{align}
        \label{eq:isoscalar_decomp_final}
        \mA_\lambda(s,t,u) &= \sum^\jmax_{j\geqslant|\lambda|} (2j+1) \; d_{\lambda0}^j(\theta_s) \; a_{j\lambda}(s) \nonumber \\
        &\qquad + \sum_{\lamp}  (-1)^{\lambda + 1}  \; d_{\lamp\lambda}^J(\omega_t)  \Bigg[ \sum^\jmax_{\jp\geqslant|\lamp|}  (2\jp+1) \; d_{\lamp0}^\jp(\theta_t) \; a_{\jp\lamp}(t) \Bigg ] \\
        &\qquad + \sum_{\lamp} (-1)^{\lamp +1} \; d_{\lamp\lambda}^J(\omega_u) \Bigg[ \sum^\jmax_{\jp\geqslant|\lamp|}  (2\jp+1) \; d_{\lamp0}^\jp(\theta_u) \; a_{\jp\lamp}(u) \Bigg ] \nonumber
    \end{align}
and project out the $j$-th partial wave:
    \begin{align}
   \label{eq:pw_as_isobar}
        &A_{j\lambda}(s) = \frac{1}{2} \int d\cos\theta^\prime \; d^j_{\lambda0}(\theta^\prime) \; \mA_\lambda(s,t(s,\theta^\prime),u(s,\theta^\prime))  \\
        &= a_{j\lambda}(s) + (-1)^{\lambda+1} \;
        \bigg[ \sum_{\lamp \jp}
        (2\jp +1) \, \int d\cos\theta^\prime \; d_{\lambda0}^j(\theta^\prime) \; d_{\lamp \lambda}^J(\omega_t^\prime) \; d_{\lamp 0}^\jp(\theta_t^\prime) \; a_{\jp \lamp}(t^\prime)
        \bigg]
        \nonumber
    \end{align}
where $t^\prime = t(s,\theta^\prime)$ and $\theta_t^\prime = \theta_t(s,\theta^\prime)$ according to Eq.~\eqref{eq:costhetat}.
We see isobars in both the direct- and cross-channels must contribute to the full partial wave. Additionally, we see Eq.~\eqref{eq:isoscalar_decomp_final} satisfies Bose symmetry---by exchanging $u$ and $t$, one gets a relative factor $(-1)^{I+\lambda}$ as Eq.~\eqref{eq:bose_iso}, with $I=1$ in this case.

Physically these isobar amplitudes may be thought of in terms of sequential two-body decays, with $a_{j\lambda}(s)$ representing the particle $X_J$ decaying with helicity-$\lambda$ into an intermediate two-pion state of spin-$j$ and a spectator pion. Crossing and Bose symmetry allow for every such intermediate state to propagate in each of the $s$-, $t$-, and $u$-channels thus motivating the expansion in Eq.~\eqref{eq:isoscalar_decomp_final}.

 Lastly, from the definition Eq.~\eqref{eq:isobar_def}, we see that the isobar partial wave amplitudes have the same kinematic singularities as the partial wave amplitudes of Eq.~\eqref{eq:isospin_pw_expansion}, as discussed in Sec.~\ref{sec:kinsincon}. This means we can define KSF isobar amplitudes exactly as in Eq.~\eqref{eq:def_ksf_partialwave}. The $t$- and $u$- channel terms in Eq.~\eqref{eq:isoscalar_decomp_final} seemingly introduce kinematic singularities in the cross-channel variables which are not present in our previous discussion of KSF partial waves. Specifically, looking at only the $t$-channel piece of Eq.~\eqref{eq:isoscalar_decomp_final} together with Eq.~\eqref{eq:H_hat}:
    \begin{align}
    \label{eq:kinematic_kernel}
        \mAhat_\lambda(s,t,u) &\propto
        \bigg[ K^{-1}_{j\lambda}(s, t) \;
        d_{\lamp\lambda}^J(\omega_t) \;
        K_{\jp\lamp}(t, s) \bigg] \;
        \hat{d}_{\lamp0}^\jp(\theta_t) \; \hat{a}_{\jp\lamp}(t)
        ~.
    \end{align}
We see the term in square brackets are the elements of the KSF matrix $\hat{\mathbb{C}}(s,t)$ in Eq.~\eqref{def:C_hat}. Similarly the $u$-channel term is proportional to the elements of $\hat{\mathbb{C}}(s,u)$. This means the kinematic singularities introduced by cross-channel isobars are taken care of by the crossing matrix such that Eq.~\eqref{eq:pw_as_isobar} still satisfies Eq.~\eqref{eq:def_ksf_partialwave}.

\paragraph{Unitarity Relations.}
Unitarity serves as a constraint on the discontinuity across the RHC of each isobar in Eq.~\eqref{eq:dispersion_relation}. Specifically, at fixed $t$ and $u$ we compute the discontinuity:
    \begin{equation}
        \Disc \mA_{\lambda}(s,t,u) = \frac{1}{2i} \bigg[ \mA_{\lambda}(s + i\epsilon,t,u) - \mA_{\lambda}(s - i\epsilon,t,u) \bigg]
    \end{equation}
by assuming a two-pion intermediate state,
    \begin{equation}
        X_J(p_X) \, \pi(p_3) \to \pi(q_1) \pi(q_2) \to \pi(p_1)\pi(p_2)
    \end{equation}
and integrating over the allowed two-body phase space. This amounts to calculating
    \begin{align}
        \label{eq:unitary_relation}
        \Disc \mA_\lambda(p_X \, p_3 \to p_1 \, p_2) &= \frac{\rho(s)}{64\pi^2} \int d\Omega^\prime_s \; {\mathcal{T}^*}^{(1)}(q_1 \, q_2 \to p_1 \, p_2) \; \mA_\lambda(p_X \, p_3 \to q_1 \, q_2) \nonumber \\
        &= \frac{\rho(s)}{64\pi^2} \int d\Omega^\prime_s \; {\mathcal{T}^*}^{(1)}(s, t^{\prime\prime}, u^{\prime\prime}) \; \mA_\lambda(s, t^\prime, u^\prime)
    \end{align}
from the isobar expanded amplitude, Eq.~\eqref{eq:isoscalar_decomp_final}, with $t^{\prime} = t(s,\theta^{\prime}_s)$, $t^{\prime\prime} = t(s,\theta^{\prime\prime}_s)$ and
    \begin{equation}
        \label{eq:pipi_amp}
        \mathcal{T}^{(I)}(s,z_s) = 32\pi \sum_{\ell = 0}^\infty (2\ell+1) \; P_\ell(\cos\theta_s) \; \tau^{(I)}_\ell(s)
    \end{equation}
is the standard decomposition of elastic pion scattering amplitude into partial waves, $\tau^{(I)}_\ell(s)$, and $\rho(s) = \sqrt{1-4m^2/s}$ is the two-body phase space function. The integration in Eq.~\eqref{eq:unitary_relation} is over the solid angle $d\Omega_s^\prime = \sin\theta^\prime_s \, d\theta^\prime_s \, d\phi^\prime$ of the intermediate state momenta related by:
    \begin{equation}
        \cos\theta_s^{\prime\prime} = \cos\theta_s \, \cos\theta_s^\prime + \cos \phi^\prime \; \sin\theta_s
        \sin\theta^\prime_s\;.
    \end{equation}

Inserting Eqs.~\eqref{eq:isoscalar_decomp_final} and \eqref{eq:pipi_amp} into Eq.~\eqref{eq:unitary_relation}, carrying out the angular integration, and taking the partial wave projection we arrive at the unitarity equation:
    \begin{align}
        \label{eq:kt_isoscalar}
        \Disc a_{j\lambda}(s) &= \rho(s) \; {\tau^*}_j^{(1)}(s) \; \times
        \\
        \bigg[ \; a_{j\lambda} (s)
        &+  (-1)^{\lambda  +1} \,\sum_{\lamp \jp}  (2\jp +1) \int d\cos\theta^\prime \; d_{\lambda0}^j(\theta^\prime) \; d_{\lamp \lambda}^J(\omega_t^\prime) \; d_{\lamp 0}^\jp(\theta_t^\prime) \; a_{\jp \lamp}(t^\prime)  \bigg] \; .
        \nonumber
    \end{align}
In terms of KSF functions, we write the unitarity relation in the compact form:
\begin{subequations}
    \label{eq:isoscalar_discontinuity_ksf}
    \begin{align}
        \label{eq:kt_ksf_isoscalar}
        \Disc \hat{a}_{j\lambda}(s) &= \rho(s) \; {\tau^*}_j^{(1)}(s) \; \bigg[ \; \hat{a}_{j\lambda} (s) + \tilde{a}_{j\lambda}(s) \bigg]
    \end{align}
where
    \begin{equation}
        \label{eq:ksf_inhomogeneity}
        \tilde{a}_{j\lambda}(s) = (-1)^{\lambda + 1} \sum_{\lamp \jp}  (2\jp +1) \int d\cos\theta^\prime \;
        \hat{C}^{j \jp}_{\lambda\lamp}(s,\theta^\prime) \;
        \hat{d}_{\lambda0}^j(\theta^\prime) \;
        \hat{d}_{\lamp 0}^\jp(\theta_t^\prime) \; \hat{a}_{\jp \lamp}(t^\prime)  \,
    \end{equation}
\end{subequations}
contains the three-body contribution from the cross-channel and is often referred to as the \textit{inhomogeneity} of the KT equation. The kernel function,
    \begin{equation}
        \label{C_kernel}
        \hat{C}_{\lambda\lamp}^{j\jp}(s,\theta) = \big(\sin^2\theta\big)^{|\lambda|} \;
        \bigg[
            K^{-1}_{j\lambda}(s,\theta)
            \; d_{\lamp \lambda}^J(\omega_t(s,\theta)) \;
            K_{\jp\lamp}(t(s,\theta), \theta_t(s,\theta))
        \bigg]
    \end{equation}
contains the appropriate combination of crossing-matrix and factored out kinematic functions such that Eq.~\eqref{eq:isoscalar_discontinuity_ksf} is free of all non-dynamical singularities except at \mbox{(pseudo-)}threshold. We note that the kernel itself is a KSF quantity in the scattering kinematics of Eq.~\eqref{eq:scatter_process}. Complications may arise in numerical solutions of the KT equations near the three threshold points identified in Sec.~\ref{sec:kinsincon} due to a necessary contour deformation to analytically continue the kinematics between the direct and cross channels as discussed in Appendix~\ref{sec:solution_strategies} and references therein.

We see the simplest case is that considered in Refs.~\cite{Niecknig:2012sj,Danilkin:2014cra} for the $J^{PC}=1^{--}$ decay, in which when truncated to $\jmax =1$ gives the only nonzero inhomogeneity kernel, $C_{11}^{11}(s,\theta) = \sin^2\theta$. Using the parity relation Eq.~\eqref{eq:helicity_parity}, the KT equations of Eq.~\eqref{eq:isoscalar_discontinuity_ksf} reduce to the equations derived by the authors:
    \begin{equation}
        \Disc \hat{a}(s) = \rho(s) \; {\tau^*}_1^{(1)}(s) \; \bigg[\hat{a} (s) + \frac{3}{2} \, \int
        d\cos\theta^\prime \sin^2\theta^\prime  \; \hat{a}(t(s,\theta^\prime)) \bigg] ~.
    \end{equation}

\subsection[Isovector decay ($a_J$ and $\pi_J$)]{\boldmath Isovector decay ($a_J$ and $\pi_J$)}
The derivation for the $I =1$ case proceeds very similar as the isoscalar case except now we must allow the possibility of intermediate states with $I = 0, \, 1,$ or $2$. In this case we write the isospin decomposition right away with Eqs.~\eqref{eq:crossing_relation_ss_t} and~\eqref{eq:crossing_relation_ss_u}:
    \begin{align}
        \mA^{(s)abcd}_\lambda(s,t,u) = \bar{\mA}^{(s)abcd}_\lambda(s,t,u) &+ \sum_\lamp (-1)^{\lambda} \; d_{\lamp\lambda}^J(\omega_t) \; \bar{\mA}_\lamp^{(s)acbd}(t,s,u)
        \nonumber \\
        &+ \sum_\lamp (-1)^{\lamp} \; d_{\lamp\lambda}^J(\omega_u) \; \bar{\mA}_\lamp^{(s)cbad}(u,t,s)
    \end{align}
and project out the $I$-th isospin projection in the $s$-channel with Eq.~\eqref{eq:isospin_amplitudes} and
with similar definitions for $\Hbar{I}{\lambda}(s,t,u)$. Then the isobar decomposition for the isovector case is:
    \begin{align}
    \label{eq:isobar_decomp_isovector}
        \mA^{(I)}_\lambda(s,t,u) = \Hbar{I}{\lambda}(s,t,u) &+ \sum_{\lamp} (-1)^\lambda \; d_{\lamp\lambda}^J(\omega_t) \; \Hbar{\Ip}{\lamp}(t,s,u) \; \frac{1}{2}C_{I\Ip}  \\
        &+ \sum_{\lamp} (-1)^\lamp \; d_{\lamp\lambda}^J(\omega_u) \; \Hbar{\Ip}{\lamp}(u,t,s) \; \frac{1}{2}C_{I\Ip} \; (-1)^{I+\Ip}
        \nonumber
    \end{align}
where the coefficients $C_{I\Ip}$ are the elements of the isospin crossing matrix:
    \begin{align}
        \frac{1}{2}C_{I\Ip} = \sum_{a,b,c,d} P_{abcd}^{(I)} \; P_{acbd}^{(\Ip)} =
        \begin{pmatrix*}[c]
        \frac{1}{3} & 1 & \frac{5}{3} \\
        \frac{1}{3} & \frac{1}{2} & -\frac{5}{6} \\
        \frac{1}{3} & -\frac{1}{2} & \frac{1}{6}
        \end{pmatrix*} \; .
    \end{align}
In terms of isobar amplitudes,
    \begin{align}
        \mA_\lambda^{(I)}(s,t,u) &= \sum_{j \geqslant |\lambda|}^\jmax (2j+1) \; d_{\lambda0}^j(\theta_s) \; \hbar{I}{j\lambda}(s)
        \nonumber \\
        &+ \sum_{\lamp \jp \Ip} \; (-1)^\lambda \; (2\jp+1) \; d_{\lamp\lambda}^J(\omega_t) \;  d_{\lamp0}^\jp(\theta_t) \; \hbar{\Ip}{\jp\lamp}(t) \; \frac{1}{2}C_{I\Ip}
        \\
        &+ \sum_{\lamp\jp \Ip} \; (-1)^\lamp (2\jp+1) \; d_{\lamp\lambda}^J(\omega_u) \;  d_{\lamp0}^\jp(\theta_u) \; \hbar{\Ip}{\jp\lamp}(u) \; \frac{1}{2}C_{I\Ip}  \; (-1)^{I+\Ip}
        \nonumber
    \end{align}
where the sums over $j$ are restricted to $j+I$ is even by Bose symmetry. We additionally see the appropriate relative factor of $(-1)^{\lambda+I}$ between the cross-channel sums.
The isospin-projected partial wave amplitude is then related to the isobar amplitudes by:
    \begin{align}
        A_{j\lambda}^{(I)}(s) &= \hbar{I}{j\lambda}(s)
         \\
        &+ (-1)^\lambda \sum_{\lamp\jp\Ip}
        \bigg[
        (2\jp+1) \, \frac{1}{2}C_{I\Ip}
        \int
        d\cos\theta^\prime \; d_{\lambda0}^j(\theta^\prime) \; d_{\lamp\lambda}^J(\omega_t^\prime) \; d_{\lamp0}^\jp(\theta_t^\prime) \;\hbar{\Ip}{\jp\lamp}(t^\prime)
        \bigg]~.
        \nonumber
    \end{align}

Now to derive the KT equations for the isoscalar case, we write the analogous  unitarity relation to Eq.~\eqref{eq:unitary_relation}:
    \begin{align}
        \label{eq:unitary_relation2}
        \Disc \mA^{(I)}_\lambda(s,t,u) =
        \frac{\rho(s)}{64\pi^2} \int d\Omega^\prime_s \; {\mathcal{T}^*}^{(I)}(s, t^{\prime\prime}, u^{\prime\prime}) \times \mA^{(I)}_\lambda(s, t^\prime, u^\prime)
    \end{align}
where the only difference is that isospin conservation requires the intermediate \(\pi\pi\) amplitude to have the same isospin projection as the total amplitude for $I = 0, 1,$ or 2. Then carrying out the angular integration and taking the partial-wave projection of Eq.~\eqref{eq:unitary_relation2}, the unitarity relation for the isovector amplitude is
    \begin{align}
        \label{eq:kt_isovector}
        &\Disc a^{(I)}_{j\lambda}(s) = \rho(s) \; {\tau^*}_j^{(I)}(s) \; \times
         \\
        &\bigg[ \; a^{(I)}_{j\lambda} (s)
        + (-1)^\lambda \sum_{\Ip \lamp \jp}
        (2\jp +1) \; \frac{1}{2}C_{I\Ip}\int d\cos\theta^\prime \; d_{\lambda0}^j(\theta^\prime) \; d_{\lamp \lambda}^J(\omega_t^\prime) \; d_{\lamp 0}^\jp(\theta_t^\prime) \; a^{(\Ip)}_{\jp \lamp}(t^\prime)  \bigg]~,
        \nonumber
    \end{align}
which, as expected, is similar to Eq.~\eqref{eq:kt_isoscalar} with an additional sum over cross-channel isospin. Finally, in terms of KSF quantities, we generalize Eq.~\eqref{eq:kt_ksf_isoscalar} as:
\begin{subequations}
    \begin{align}
        \label{eq:kt_ksf_isovector}
        \Disc \hat{a}^{(I)}_{j\lambda}(s) &= \rho(s) \; {\tau^*}_j^{(I)}(s) \bigg[ \; \hat{a}^{(I)}_{j\lambda} (s) +  \tilde{a}^{(I)}_{j\lambda}(s) \bigg]
    \end{align}
for the inhomogeneity for isospin $I$ given by
    \begin{gather}
        \label{eq:ksf_inhomogeneity_isovector}
        \tilde{a}^{(I)}_{j\lambda}(s) = (-1)^\lambda \sum_{\Ip \lamp \jp}
        (2\jp +1) \, \frac{1}{2}C_{I\Ip} \int d\cos\theta^\prime \;
        \hat{C}^{j \jp}_{\lambda\lamp}(s,\theta^\prime) \;
        \hat{d}_{\lambda0}^j(\theta^\prime) \;
        \hat{d}_{\lamp 0}^\jp(\theta^\prime_t) \; \hat{a}^{(\Ip)}_{\jp \lamp}(t^\prime)  \, ,
    \end{gather}
\end{subequations}
where the kernel function is independent of isospin and is the same as Eq.~\eqref{C_kernel}.

We note Eqs.~\eqref{eq:kt_ksf_isovector} and~\eqref{eq:ksf_inhomogeneity_isovector} reduce to the unitarity equations for the $\pi\pi$ scattering case derived in Ref. \cite{Albaladejo:2018gif} when $J=0$. Additionally a similar isobar decomposition is considered in Ref.~\cite{Brehm1981} in the context of the $a_1 \to 3\pi$ decay.

\section{Summary and outlook}
The KT formalism has recently acquired great importance, and it has been applied to study many reactions. In view of this, in this manuscript we have generalized it to make it usable in the decays of particles with arbitrary spin, parity, and charge conjugation. To perform the KT decomposition in a proper way, the crossing symmetry relation of the helicity amplitudes has been carefully taken into account. By analyzing the crossing matrix, we have also been able to derive reduced helicity isobars which are free of kinematic singularities, as well as the kinematic constraints that they must satisfied.

Because of its more immediate interest, we have focused on the cases of isoscalar and isovector particles decaying into three pions. The formalism presented can be readily applied to study important processes, such as the decays $a_1 \to 3\pi$, or similar ones. The generalization to decays of particles with $I=2,3$ should be quite straightforward. Indeed, the key step of our construction of KT equations is incorporating in the decomposition crossing relations, which are easily generalized to other isospin values.

We also note that the formalism presented here does not apply only to strong, isospin-conserving processes. It can be readily applied to other non-isospin-conserving (either strong or electroweak) reactions, where the amplitudes are decomposed into isospin components by means of Wigner-Eckart theorem (see {\it e.g.} Ref.~\cite{Albaladejo:2017hhj}).

\acknowledgments
This work was supported by the U.S.~Department of Energy under Grants
No.~DE-AC05-06OR23177 
and No.~DE-FG02-87ER40365, 
the U.S.~National Science Foundation under Grant 
No.~PHY-1415459, 
by the Ministerio de Ciencia, Innovaci\'on y Universidades (Spain) under Grants No.~FPA2016-77313-P and No.~FPA2016-75654-C2-2-P, 
by  PAPIIT-DGAPA (UNAM, Mexico) under Grant No.~IA101819, 
and CONACYT (Mexico) under Grants No.~251817,  
No.~734789 
and No.~A1-S-21389, 
by the DFG [Projektnummer 204404729 - SFB 1044] and in part through the Cluster of Excellence (PRISMA+ EXC
2118/1) within the German Excellence Strategy (Project ID 39083149). 
V.M. is supported by the Comunidad Aut\'onoma de Madrid through the Programa de Atracci\'on de Talento Investigador 2018 (Modalidad 1). 

\appendix
\section[Derivation of crossing relation]{\boldmath Derivation of crossing relation}\label{app:proofcrossfull}
In this Appendix we present a general derivation of Eq.~\eqref{eq:crossfull} which relates the helicity amplitudes defined in the different scattering frames by the permutation of Mandelstam variables,
\begin{equation}\label{eq:crossfullApp}
    \mA_\lambda^{(t)acbd}(t,s,u) = (-1)^\lambda \; \mA_\lambda^{(s)acbd}(t,s,u)~.
\end{equation}
We recall that the amplitudes are defined by the representations of the four-momenta and polarization vectors in their respective frame as in Eqs.~\eqref{eq:hel_amp_s}~and~\eqref{eq:hel_amp_t}. We make use of the expressions of the momenta of the particles expressed in the $s$- and $t$-channel frames, Eqs.~\eqref{eq:kin_s}~and~\eqref{eq:kin_t} to first relate the arguments of the r.h.s. in Eq.~\eqref{eq:crossfullApp}, in each frame. The usual polarization vectors have components [see Eq.~\eqref{eq:pol_vec_J1}] which satisfy:
\begin{align}
    {\epsilon'}^{(t)}_\pm(s,t,u) & = \epsilon^{(s)}_\pm(t,u,s)~, \mand
    {\epsilon'}^{(t)}_0(s,t,u) = \epsilon^{(s)}_0(t,u,s)~,
\end{align}
and, since the polarization tensor of any order $J > 1$ can be built from the above polarization vectors (see {\it e.g.} Eq.~\eqref{eq:pol_vec_genJ} for $J=2$), the relations above are more generally:
\begin{equation}
\label{pol_cross}
    {\epsilon^\prime}^{(t)}_\lambda(s,t,u) = \epsilon^{(s)}_\lambda(t,u,s)~.
\end{equation}
Now, we relate the momenta $p'_i(s,t,u)$ (see Fig.~\ref{fig:stchannel}) with the analogous momenta $p_i(t,u,s)$ in the $s$-channel. For the initial state pion, $p'_1(s,t,u)$, we find:
\begin{equation}
\label{incoming_cross}
    p'_1(s,t,u) = \left( - E_\pi(t), \, p(t) \; \hat{z} \right) = -p_3(t,u,s)~.
\end{equation}
To relate the final state momenta, $p'_{2,3}(s,t,u)$, in the $t$-channel with the four-vectors $p_{2,1}(t,u,s)$ in the $s$-channel, we must first relate $\hat{q}'_t(s,t,u)$ with $\hat{q}_s(t,u,s)$. From their definitions, Eqs.~\eqref{eq:vector_q_s},~\eqref{eq:vector_q_t}, and~\eqref{eqs:sins_etas}:
\begin{align}
\label{outpoing_cross}
    \hat{q}'_t(s,t,u) 
    & = - \sin\theta_t(s,t,u) \; \hat{x} + \cos\theta_t(s,t,u) \; \hat{z} \nonumber \\
    & = - \sin\theta_s(t,u,s) \; \hat{x} - \cos\theta_s(t,u,s) \; \hat{z} = - \hat{q}_s(t,u,s)~,
\end{align}
and therefore the momenta satisfy:
\begin{subequations}
\begin{align}
    p'_3(s,t,u) & = - \left( \frac{\sqrt{t}}{2} , q(t) \; \hat{q}'_t(s,t,u) \right) = - \left( \frac{\sqrt{t}}{2} , -q(t) \; \hat{q}_s(t,u,s) \right) = - p_2(t,u,s)~,
\end{align}
and
\begin{align}
    p'_2(s,t,u) & = \left( \frac{\sqrt{t}}{2} ,-q(t)  \; \hat{q}'_t(s,t,u) \right) = \left( \frac{\sqrt{t}}{2} , +q(t) \; \hat{q}_s(t,u,s) \right) =   p_1(t,u,s)~.
\end{align}
\end{subequations}

Combining Eqs.~\eqref{pol_cross},~\eqref{incoming_cross}, and~\eqref{outpoing_cross}, with the definitions Eq.~\eqref{eq:hel_amp_s} and~\eqref{eq:hel_amp_t}, we have:
\begin{align}
    \mA^{(t)acbd}_{\lambda}(t,s,u) 
& = \mA^{abcd}(\epsilon'^{(t)}_\lambda(s,t,u), \, p'_1(s,t,u), \, p'_2(s,t,u), \, p'_3(s,t,u)) \nonumber\\
& = \mA^{abcd}(\epsilon^{(s)}_\lambda(t,u,s), \, -p_3(t,u,s), \, +p_1(t,u,s), \, -p_2(t,u,s))  \nonumber\\
& = \mA^{acbd}(\epsilon^{(s)}_\lambda(t,u,s), \, +p_2(t,u,s), \, +p_1(t,u,s), \, +p_3(t,u,s))  \nonumber\\
& = \mA^{acdb}(\epsilon^{(s)}_\lambda(t,u,s), \, +p_1(t,u,s), \, +p_2(t,u,s), \, +p_3(t,u,s))  \nonumber\\
& = \mA_\lambda^{(s)acdb}(t,u,s) = (-1)^\lambda \mA_\lambda^{(s)acbd}(t,s,u)~,
\end{align}
which is the result in Eq.~\eqref{eq:crossfull}. In the intermediate steps above we have used the additional result regarding the interchange of pion isospin indices:
\begin{equation}
    \mA^{abcd}(\epsilon,p_1,p_2,p_3) = \mA^{acbd}(\epsilon,-p_3,p_2,-p_1) = \mA^{abdc}(\epsilon,p_2,p_1,p_3)~.
\end{equation}

\section{Proof of crossing symmetry of KT decomposition}\label{app:proof2}
The KT formalism aims to restore the crossing symmetry of the full amplitude using a decomposition into a truncated sum of isobar amplitudes. As such we have payed particular attention to the crossing relations between helicity amplitudes in different scattering frames in Sec.~\ref{sec:crossing}. The resulting isobar decomposition in Eq.~\eqref{eq:KT_isoscalar} presents the combination of truncated isobar sums in each channel: $\bar{\mA}_{\lambda}(s,t,u)$, $\bar{\mA}_{\lambda}(t,s,u)$, $\bar{\mA}_{\lambda}(u,t,s)$ that satisfies the crossing symmetry relation in Eq.~\eqref{eq:crossfull_isoscalar}. This appendix provides an explicit proof of this fact.

We start by inserting the definition \eqref{eq:KT_isoscalar} into the r.h.s. of Eq.~\eqref{eq:crossfull_isoscalar}, using $\omega_t(s,t,u) = \omega_t(t,s,u)$:
\begin{align}
\sum_{\lamp} d_{\lamp\lambda}^J(\omega_t) \;& \mA_\lamp(t,s,u) = (-1)^{1+\lambda} \sum_{\lambda'} d^J_{\lambda'\lambda}(\omega_t(s,t,u))  \; \bar{\mA}_{\lambda'}(t,s,u) \nonumber \\
& + \sum_{\lambda''} \sum_{\lambda'} (-1)^{\lambda'+\lambda} \;  d^J_{\lambda''\lambda'}(\omega_t(s,t,u)) \; d^J_{\lambda'\lambda}(\omega_t(s,t,u))\; \bar{\mA}_{\lambda''}(s,t,u) \label{eq:caux}\\
& + \sum_{\lambda''} \sum_{\lambda'} (-1)^{\lambda''+\lambda} \; d^J_{\lambda''\lambda'}(\omega_u(t,s,u)) \; d^J_{\lambda'\lambda}(\omega_t(s,t,u)) \; \bar{\mA}_{\lambda''}(u,s,t)~. \nonumber
\end{align}
The first term in Eq.~\eqref{eq:caux} is clearly equal to the second term in the definition of $\mA_{\lambda}(s,t,u)$ in Eq.~\eqref{eq:KT_isoscalar}. To identify the second term in Eq.~\eqref{eq:caux}, we perform the $\lambda'$ summation, obtaining:
\begin{equation}
\sum_{\lambda'}  (-1)^{\lambda'} \, d^J_{\lambda''\lambda'}(\omega_t) \;  d^J_{\lambda'\lambda}(\omega_t) = 
(-1)^{\lambda''} \; \delta_{\lambda\lambda''}~.
\end{equation}
Inserting this into the second term of Eq.~\eqref{eq:caux}, we get:
\begin{align}
\sum_{\lambda''} \sum_{\lambda'} (-1)^{\lambda'+\lambda} \; &d^J_{\lambda''\lambda'}(\omega_t(s,t,u)) \;  d^J_{\lambda'\lambda}(\omega_t(s,t,u)) \; \bar{\mA}_{\lambda''}(s,t,u) ) \nonumber \\
& = \sum_{\lambda''} (-1)^{\lambda'' + \lambda} \; \delta_{\lambda\lambda''} \; \bar{\mA}_{\lambda''}(s,t,u) = \bar{\mA}_{\lambda}(s,t,u)~,
\end{align}
which we identify as the first time of $\mA_{\lambda}(s,t,u)$ in Eq.~\eqref{eq:KT_isoscalar}. Finally to match the third term we note that $\omega_u(t,s,u) = \omega_t(t,u,s)$ and the $\lambda'$ summation becomes:
\begin{equation}
\label{third_term}
\sum_{\lambda'} d^J_{\lambda'\lambda}(\omega_t(s,t,u)) \; d^J_{\lambda''\lambda'}(\omega_t(t,u,s)) = d^J_{\lambda''\lambda}(\omega_t(s,t,u) + \omega_t(t,u,s))~.
\end{equation}
From definitions Eq.~\eqref{eq:omega}, we see
\begin{subequations}\begin{align}
\cos\left( \omega_t(s,t,u) + \omega_t(t,u,s) \right) & = \hphantom{+}\cos\left( \omega_t(s,u,t) \right)~,
\end{align}
and
\begin{align}
\sin\left( \omega_t(s,t,u) + \omega_t(t,u,s) \right) & =           - \sin\left( \omega_t(s,u,t) \right)~,
\end{align}
\end{subequations}
which imply the argument of the single Wigner function in Eq.~\eqref{third_term} is: 
\begin{equation}
\label{omega_tu_rel}
\omega_t(s,t,u) + \omega_t(t,u,s) = - \omega_t(s,u,t) = - \omega_u(s,t,u)~.
\end{equation}
Therefore, inserting Eq.~\eqref{omega_tu_rel} to Eq.~\eqref{third_term} and using the Bose symmetry relation of Eq.~\eqref{eq:bose_iso}, the third term in Eq.~\eqref{eq:caux} becomes:
\begin{align}
\label{last_term}
\sum_{\lambda''} (-1)^{\lambda + \lambda''} \;  d^J_{\lambda'' \lambda}(-\omega_u(s,t,u)) \;
\bar{\mA}_{\lambda''} (u,s,t)
= \sum_{\lambda''}  d^J_{\lambda'' \lambda}(\omega_u(s,t,u)) \;
\bar{\mA}_{\lambda''}(u,s,t) &
\\
 = \sum_{\lambda'} (-1)^{1+\lambda'} \; d^J_{\lambda'\lambda}(\omega_u(s,t,u)) \;
\bar{\mA}_{\lambda'}(u,t,s) &~, \nonumber
\end{align}
which corresponds to the final remaining term in the definition of $\mA_{\lambda}(s,t,u)$ in Eq.~\eqref{eq:KT_isoscalar}.

\section{Relation between helicity amplitudes and invariant functions}\label{app:examples}
In this appendix we study the relation between invariant and helicity amplitudes. This is done for particular cases of $X_J$ quantum numbers. The purpose of the appendix is twofold: first, to show how to compute the invariant amplitudes from the helicity amplitudes that we have used through the manuscript; and second, to give an example of how the crossing relations,  Eqs.~\eqref{eq:crossing_relation}~and~\eqref{eq:crossfull}, emerge from the definitions of the amplitudes.
\subsection[Axial vector meson, $a_1: I^G(J^{PC})= 1^-(1^{++})$]{Axial vector meson,  \boldmath $a_1$ : $I^G(J^{PC})= 1^-(1^{++})$}
\label{app:a1_details}

We start by considering the specific process
\begin{equation}
    a_1^a(\epsilon,\, p_X) \; \pi^b(p_3) \to \pi^c(p_1) \; \pi^d(p_2)~,
\end{equation}
which is also the subject of Sec.~\ref{sec:singul_a1}.
As before, the amplitude is denoted $\mA^{abcd}(\epsilon(p_X),p_3;p_1,p_2)$ [\textit{cf.} Eq.~\eqref{master_amp}] but instead of the decomposition into helicity partial wave amplitudes as in Eq.~\eqref{eq:hel_amp_s}, we write the amplitude in terms of the most general Lorentz covariant structure satisfying parity and Bose symmetry (\textit{c.f.} Eqs.~\eqref{eq:helicity_parity} and~\eqref{eq:bose_iso}) as:
\begin{align}
\label{cov_rep}
\mA^{abcd}(\epsilon(p_{a_1}), p_3; p_1,p_2) & =
\epsilon \cdot
(p_1 + p_2)\, F^{abcd} \left( s,t,u \right) +
\epsilon \cdot
(p_1 - p_2)\, G^{abcd} \left( s,t,u \right)~.
\end{align}
We see the functions $F$ and $G$ are Lorentz scalars and are thus referred to as invariant amplitudes.

The components of the usual rank-one polarization vector, are
\begin{align}
    \label{eq:pol_vec_J1}
    \epsilon_0  = \left( \frac{\modk }{M}, \; \frac{E_X}{M} \hat{z} \right)~,
    \mand
    \epsilon_+  = \left( 0, \; \frac{\hat{x} + i \hat{y}}{\sqrt{2}} \right)~,
\end{align}
for helicity projection, $\lambda = 0, \, +1$ respectively. We see in the covariant representation of Eq.~\eqref{cov_rep}, the helicity amplitude is dictated by the form of the polarization vector.
From this we may derive relations between the $s$- and $t$-channel helicity amplitudes and the invariant amplitudes. For the $s$-channel, from Eq.~\eqref{eq:hel_amp_s}
    \begin{align}\label{eq:Hs_nocross}
        \begin{pmatrix}
            \sqrt{2} \; \mA_+^{(s)abcd}(s,t,u) \\
            \mA_0^{(s)abcd}(s,t,u)
        \end{pmatrix}
         =
         \frac{1}{M}
        \begin{pmatrix}
         0 & 2\, M \, q(s) \sin \theta_s \\
        p(s)\sqrt{s} & \quad  - 2 \, q(s) \, E_{a_1}(s) \cos\theta_s
    \end{pmatrix}
&
    \begin{pmatrix}
        F^{abcd}(s,t,u) \\
        G^{abcd}(s,t,u)
        \end{pmatrix} \nonumber \\
 \equiv
        \mathbb{Q}(s,t)\,
        &\begin{pmatrix}
        F^{abcd}(s,t,u) \\
        G^{abcd}(s,t,u)
        \end{pmatrix}
    \end{align}
where the factor $\sqrt{2}$ in front of $\mA_+^{(s)abcd}(s,t,u)$ and kinematic matrix $\mathbb{Q}(s,t)$ are introduced for later convenience.

Using Eq.~\eqref{eq:hel_amp_t} we may derive the analogous equation for the $t$-channel with appropriate changes of momenta and isospin indices by enforcing
\begin{equation}
    \label{master_amp_Relatioon}
\mA^{abcd}(\epsilon, p_3, p_1,p_2,) =
\mA^{acbd}(\epsilon, -p_1, -p_3,p_2)~.
\end{equation}
From Eq.~\eqref{cov_rep}, we notice:
\begin{subequations}
    \label{mmatrix}
    \begin{equation}
    \left(\begin{array}{cc} \epsilon \cdot (-p_3 + p_2) & \epsilon \cdot (-p_3 - p_2) \end{array}\right) =
     \mathbb{M} \;
    \left(\begin{array}{cc} \epsilon \cdot (p_1 + p_2) & \epsilon \cdot (p_1 - p_2) \end{array}\right)~.
    \end{equation}
where the constant matrix:
    \begin{equation}
        \mathbb{M} = -\frac{1}{2}
        \begin{pmatrix}
        1 & 3 \\
        1 & -1
        \end{pmatrix}~,
        \quad \text{with} \quad
        \mathbb{M}^{-1} = \mathbb{M}~,
    \end{equation}
\end{subequations}
connects the covariant kinematic factors in the different channels. Combining Eq.~\eqref{master_amp_Relatioon} and~\eqref{mmatrix}, we find invariance under isospin rotations requires the invariant amplitudes to satisfy:
    \begin{equation}
    \label{m_crossing}
        \begin{pmatrix}
        F^{abcd}(s,t,u) \\
        G^{abcd}(s,t,u)
        \end{pmatrix}
        =
        \mathbb{M}
        \begin{pmatrix}
        F^{acbd}(t,s,u) \\
        G^{acbd}(t,s,u)
        \end{pmatrix}~.
    \end{equation}
Now we find the write the analogous expression to Eq.~\eqref{eq:Hs_nocross} for the $t$-channel:
    \label{eq:Ht_nocross}
    \begin{align}
        \begin{pmatrix}
            \sqrt{2} \; \mA_+^{(t)acbd}(t,s,u) \\
            \mA_0^{(t)acbd}(t,s,u)
        \end{pmatrix}
        & =
         \diag(-1, \, 1) \;
        \mathbb{Q}(t,s)
        \begin{pmatrix}
        F^{acbd}(t,s,u) \\
        G^{acbd}(t,s,u)
        \end{pmatrix}
    \end{align}
Combining Eqs.~\eqref{eq:Hs_nocross} and~\eqref{eq:Ht_nocross} with Eq.~\eqref{m_crossing}, we may equate helicity amplitudes in different channels up to a combination of kinematic matrices:
    \begin{align}
        \label{ff}
        \begin{pmatrix}
        \sqrt{2} \; \mA_+^{(s)abcd}(s,t,u) \\
        \mA_0^{(s)abcd}(s,t,u)
        \end{pmatrix}
        =
        \mathbb{Q}(s,t) \; \mathbb{M} \; \mathbb{Q}^{-1}(t,s)
        \begin{pmatrix}
        \sqrt{2} \; \mA_+^{(t)acbd}(t,s,u) \\
        \mA_0^{(t)acbd}(t,s,u)
        \end{pmatrix}
    \end{align}
Calculating the matrix explicitly and using definitions in Eq.~\eqref{eq:omega}, we find
    \begin{equation}
        \mathbb{Q}(s,t) \; \mathbb{M} \; \mathbb{Q}^{-1}(t,s) =
        \begin{pmatrix}
         \cos \omega & -\sin \omega \\
          \sin \omega & \cos \omega
        \end{pmatrix}
    \end{equation}
which is exactly the crossing relation Eq.~\eqref{eq:crossing_relation} for $J^{PC}=1^{++}$ (\textit{c.f.} Eq.~\eqref{eq:matrix-helicity_a1} for the explicit evaluation of $d$-matrix elements for this case).
We note that taking Eq.~\eqref{eq:Ht_nocross} with the interchange $s \leftrightarrow t$ and Eq.~\eqref{eq:Hs_nocross}, we arrive at:
\begin{align}
    \label{ff2}
        \begin{pmatrix}
           - \sqrt{2} \; \mA_+^{(t)acbd}(s,t,u) \\
            \mA_0^{(t)acbd}(s,t,u)
        \end{pmatrix}
         =
        \mathbb{Q}(s,t)
        \begin{pmatrix}
        F^{acbd}(s,t,u) \\
        G^{acbd}(s,t,u)
        \end{pmatrix}
        =
     \begin{pmatrix}
        \sqrt{2} \; \mA_+^{(s)acbd}(s,t,u) \\
        \mA_0^{(s)acbd}(s,t,u)
    \end{pmatrix}
\end{align}
which is the relation between $s$- and $t$-channel kinematics with the interchange of energy variables in Eq.~\eqref{eq:crossfull}.

\subsection[Tensor meson, $a_2: I^G(J^{PC})= 1^-(2^{++})$]{\boldmath Tensor meson, $a_2$: $I^G(J^{PC})= 1^-(2^{++})$}

Similar to the above derivations we may consider the process
\begin{equation}
    a_2^a(\epsilon,\, p_{a_2}) \; \pi^b(p_3) \to \pi^c(p_1) \; \pi^d(p_2)~,
\end{equation} 
to illustrate how things change when considering the decay a natural parity meson and higher spin.

To build the general most covariant structure and identify the invariant amplitudes we first construct the rank-two polarization tensor for the $J=2$ meson from Eq.~\eqref{eq:pol_vec_J1} as:
\begin{equation} 
    \label{eq:pol_vec_genJ}
    \epsilon_{\mu\nu}^{(\lambda)} = \sum_{\lambda', \, \lambda''} 
    \braketno{1 \, \lambda'\, ; 1 \,\lambda''}{2 \,\lambda} \; \epsilon_{\mu}^{(\lambda')}
    \; \epsilon_{\nu}^{(\lambda'')}~.
\end{equation}
Explicitly this gives us for $\lambda = 0, \, +1, \, +2$:
\begin{align}
\label{eqs:pol_tens}
    \epsilon^{(2)}_{\mu\nu} & =  \epsilon_\mu^{(+)}\epsilon_\nu^{(+)}~, \nonumber \\
    \epsilon^{(1)}_{\mu\nu} & = \frac{\epsilon_\mu^{(+)}\epsilon_\nu^{(0)} + \epsilon_\mu^{(0)}\epsilon_\nu^{(+)}}{\sqrt{2}}~,\\
    \epsilon^{(0)}_{\mu\nu} & = \frac{\epsilon_\mu^{(+)}\epsilon_\nu^{(-)} + \epsilon_\mu^{(-)}\epsilon_\nu^{(+)} -2 \epsilon_\mu^{(0)}\epsilon_\nu^{(0)}}{\sqrt{6}}~. \nonumber
\end{align}
The decomposition of Eq.~\eqref{eq:pol_vec_genJ}, may be used as a template to define the polarization tensor or arbitrary rank.

With \cref{eqs:pol_tens}, we write the ``master amplitude" as a covariant expression as in Eq.~\eqref{cov_rep}:
\begin{align}
    \label{cov_rep2}
    \mA^{abcd}(\epsilon(p_{a_2}),p_3,p_1,p_2) &= 
    - \, i \,\sqrt{2} \; \epsilon_{\mu\nu} \;  K^{\mu} \; \times 
    \nonumber \\
    &\left( 
    (p_1 + p_2)^\nu \; B^{abcd}(s,t,u) +
    (p_1 - p_2)^\nu \; C^{abcd}(s,t,u)
    \right)~,
\end{align}
where $K$ is the kibble vector defined in Eq.~\eqref{eq:Kibble_s} and we take $\epsilon_{0123}= +1$. As before, the Lorentz scalar functions $B$ and $C$ are the invariant amplitudes for this process. There are only two independent invariant functions because due to the parity relation in Eq.~\eqref{eq:helicity_parity} which enforces the amplitude vanishes for $\lambda = 0$. 

Now we use Eq.~\eqref{cov_rep2} to relate the $s$-channel helicity amplitudes of Eq.~\eqref{eq:hel_amp_s} (for $\lambda = 1, \, 2$)  to the invariant functions:
    \begin{align}
    \label{hs_2}
        \begin{pmatrix}
         \mA_{2}^{(s)abcd}(s,t,u) 
         \\ \mA_{1}^{(s)abcd}(s,t,u) 
        \end{pmatrix}
        =
        \mathbb{Q}_2(s,t) 
        \begin{pmatrix}
         B^{abcd}(s,t,u) 
         \\ C^{abcd}(s,t,u) 
        \end{pmatrix}
    \end{align}
where by explicitly doing the Lorentz contraction one may show the kinematic matrix is related to the matrix considered in Eq.~\eqref{eq:Hs_nocross} by: 
    \begin{equation}
        \mathbb{Q}_2(s,t) = \frac{\sqrt{s} \, p(s) \, q(s) \; \sin\theta_s}{\sqrt{2}} \; \mathbb{Q}(s,t) = \sqrt{\frac{1}{2} \left(\frac{\phi}{4}\right)} \; \mathbb{Q}(s,t)~.
    \end{equation}
As before, we may also write the analogous expression for the $t$-channel by interchanging $p_1 \to -p_3$ and $b \leftrightarrow c$ in Eq.~\eqref{cov_rep2}. Carrying out the Lorentz contractions one will find:
    \begin{align}
        \label{ht_2}
        \begin{pmatrix}
         \mA_{2}^{(t)acbd}(t,s,u) 
         \\ \mA_{1}^{(t)acbd}(t,s,u) 
        \end{pmatrix}
        =
        \diag(1, -1) \;
        \mathbb{Q}_2(t,s) 
        \begin{pmatrix}
         B^{acbd}(t,s,u) 
         \\ C^{acbd}(t,s,u) 
        \end{pmatrix}~.
    \end{align}

 Since we have identical isospin structure as in the vector case, the invariant amplitudes in this case can be shown to obey the same relations as in Eq.~\eqref{m_crossing}:
\begin{equation}
\label{m_crossing2}
    \begin{pmatrix}
    B^{abcd}(s,t,u) \\
    C^{abcd}(s,t,u)
    \end{pmatrix}
    = 
    \mathbb{M}
    \begin{pmatrix}
    B^{acbd}(t,s,u) \\
    C^{acbd}(t,s,u)
    \end{pmatrix}~.
\end{equation}
for the same constant matrix $\mathbb{M}$ in \cref{mmatrix}.
Combining Eqs.~\eqref{hs_2},~\eqref{ht_2}, and~\eqref{m_crossing2} we arrive at the crossing relations for helicity amplitudes for the $a_2$:
    \begin{align}
        \begin{pmatrix}
         \mA_{2}^{(s)abcd}(s,t,u) 
         \\ \mA_{1}^{(s)abcd}(s,t,u) 
        \end{pmatrix}
        =
        \diag(1, -1) \;
        \mathbb{Q}_2(s,t) \; \mathbb{M} \; \mathbb{Q}^{-1}_2(t,s)
        \begin{pmatrix}
         \mA_{2}^{(t)acbd}(t,s,u) 
         \\ \mA_{1}^{(t)acbd}(t,s,u) 
        \end{pmatrix}
    \end{align}
and 
    \begin{align}
        \begin{pmatrix}
         \mA_{2}^{(t)acbd}(t,s,u) 
         \\ -\mA_{1}^{(t)abcd}(t,s,u) 
        \end{pmatrix}
        =
        \mathbb{Q}_2(s,t) \;    
        \begin{pmatrix}
        B^{acbd}(t,s,u) \\
        C^{acbd}(t,s,u)
        \end{pmatrix}
        =
        \begin{pmatrix}
         \mA_{2}^{(s)acbd}(t,s,u) 
         \\ \mA_{1}^{(s)acbd}(t,s,u) ~. 
        \end{pmatrix}
    \end{align}
Which are the analogous relations for Eqs.~\eqref{ff} and~\eqref{ff2} for $J=2$. In fact, solving for the elements of the matrix above, we see:
    \begin{equation}
        \diag(1, -1) \;
        \mathbb{Q}_2(s,t) \; \mathbb{M} \; \mathbb{Q}^{-1}_2(t,s) =        
        \begin{pmatrix}
         \cos \omega & -\sin \omega \\
          \sin \omega & \cos \omega
        \end{pmatrix}
    \end{equation}
which is identical to the crossing relation we found above a fact that can also be verified by explicitly evaluating the elements of the $d$-matrix in \cref{eq:crossing_relation}.
\section{Solution strategies}\label{sec:solution_strategies}

In the main body of the paper we introduced a systematic construction of amplitudes based on analyticity, crossing symmetry, and subchannel unitarity based on the KT formalism. In order to further make this manuscript as self-contained as possible we include a schematic for finding numerical solutions to the above KT equations.

 For simplicity, let us consider a single reduced partial wave, for a single helicity value, that we denote by $\hat{a}$. The generalization to include other reduced partial waves is straightforward. Furthermore, in what respects the numerical solution, the inclusion of coupled channels in the unitarity relations does not bring forth additional complications~\cite{Albaladejo:2017hhj}. With these simplifications in mind, Eq.~\eqref{eq:kt_ksf_isoscalar} reads:
\begin{equation}\label{eq:sols:unitarity}
    \text{Disc}\, \hat{a}(s) = \rho(s) \; \tau^\ast(s) \left( \hat{a}(s) + \widetilde{a}(s) \right)~,
\end{equation}
where the $\pi\pi$ scattering amplitude $\tau(s)$ is given, and the inhomogeneity $\widetilde{a}(s)$ is given in terms of $\hat{a}(s)$ as an angular integral, so we generically write Eq.~\eqref{eq:ksf_inhomogeneity} as:
\begin{equation}\label{eq:sols:ft}
    \widetilde{a}(s) = \int_{-1}^{+1} \mathrm{d}z\, C(s,z)\, \hat{a}\left(t(s,z) \right) = \int_{t_-(s)}^{t_+(s)} dt\, K(s,t) \; \hat{a}(t)~.
\end{equation}
where the kernels $C(s,z)$ and $K$ [$K$ is defined such that $C(s,z) = K(s,t(s,z))$] can be known, for each different case, from the analyses in the previous sections. The dependence on $s$ of the endpoints arises because of the $z$ to $t$ variable transformation in the integral, and the relation of $t$ and $z$ is given by [{\it cf.} Eq.~\eqref{eq:costhetas}]:
\begin{equation}
    t(s,z) = \frac{3m^2 + M^2 - s}{2} + 2 \; p(s) \, q(s) z~,
\end{equation}
so that the endpoints are given by:
\begin{equation}\label{eq:tminmax}
    t_\pm(s) = t(s,\pm 1) = \frac{3m^2 + M^2 - s}{2} \pm  2 \; p(s) \, q(s)~.
\end{equation}

An $n$-subtracted dispersion relation\footnote{For simplicity, the subtractions, if any, are taken at $s=0$.} of Eq.~\eqref{eq:sols:unitarity} gives:
\begin{equation}\label{eq:sols:f_no}
\hat{a}(s) = \widetilde{P}_{n-1}(s) + \widetilde{I}_{n-1}(s)~,
\end{equation}
where $\widetilde{P}_{n-1}(s)$ is a $n-1$ degree polynomial (with the understanding that $P_{i < 0} = 0$), and the dispersive integral $\widetilde{I}_{n-1}(s)$ is given by:
\begin{equation}
\widetilde{I}_{n-1}(s) = \frac{s^n}{\pi}\int_{s_\text{th}}^{\infty} \mathrm{d}s' \frac{\text{Disc}\, \hat{a}(s')}{{s'}^n (s'-s)}~.
\end{equation}
If, for a moment, we ignore the inhomogeneity $\widetilde{a}(s)$, then Eq.~\eqref{eq:sols:unitarity} becomes a standard Omn\'es problem, and the solution for $\hat{a}(s)$ is simply the Omn\'es function, $\Omega(s)$, times a polynomial. The Omn\'es function satisfies in turn the following unitarity-like relation,
\begin{equation}
    \text{disc}\, \Omega(s) = \rho(s) \; \tau^\ast(s) \; \Omega(s)~.
\end{equation}
For our problem of elastic $\pi\pi$ rescattering at hands, $\Omega(s)$ is given by:
\begin{equation}
\Omega(s) = \exp\left( \frac{s}{\pi} \int_{s_\text{th}}^\infty \mathrm{d}s' \frac{\delta(s')}{s'(s'-s)} \right)~,
\end{equation}
where $\delta(s')$ is the $\pi\pi$ scattering phase shift associated to the amplitude $t(s)$. Reintroducing in our considerations the inhomogeneity $\widetilde{f}(s)$, the most general solution of Eq.~\eqref{eq:sols:unitarity} is given by:
\begin{equation}\label{eq:sols:f_o}
    \hat{a}(s) = \Omega(s) \left( P_n(s) + I_n(s) \right)~,
\end{equation}
where $P_n(s)$ and $I_n(s)$ are hereafter referred to as polynomial and dispersive terms, and $I_n(s)$ is given by:
\begin{equation}\label{eq:sols:f_o_disp}
    I_n(s) = \frac{s^n}{\pi} \int_{4m^2}^{\infty} \mathrm{d} s' \frac{\sin\delta(s') \, \widetilde{a}(s')}{\left\lvert \Omega(s') \right\rvert {s'}^n (s'-s)}
\end{equation}

The function $a(s)$ is computed through Eqs.~\eqref{eq:sols:f_no}~and/or~\eqref{eq:sols:f_o}, and $\widetilde{a}(s)$ is computed through \eqref{eq:sols:ft} [or, as shown below, Eq.~\eqref{eq:sols:ft_ins}], and thus these equations constitute a closed system of linear integral equations for $a(s)$ and $\hat{a}(s)$. Experience shows that, because of its linearity, iteration is a good method to numerically solve this system, since convergence is usually found after five to ten iterations. Alternatively, an approach based on the matrix inversion of the discretized integral system of equations can also be exploited \cite{Niecknig:2015ija}.

\begin{figure}\centering
\input{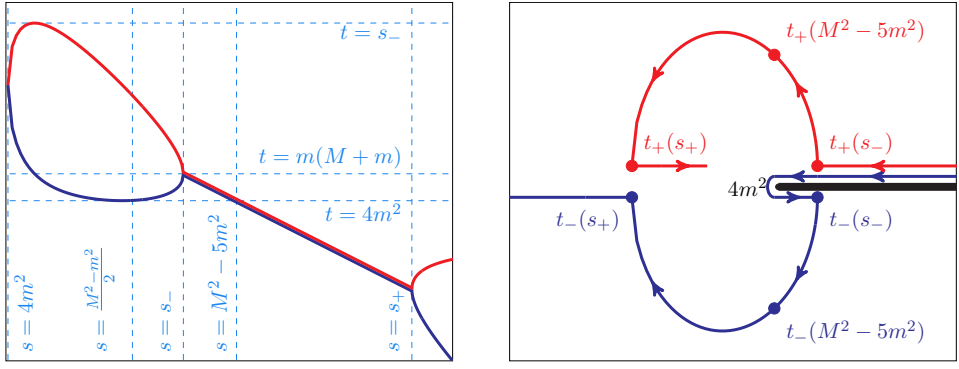}
\caption{Dependence on $s$ of $t_\pm(s)$ [Eq.~\eqref{eq:tminmax}], the endpoints of the integral in Eq.~\eqref{eq:sols:ft}. The red (blue) line curves correspond to $t_+(s)$ and $t_(s)$. Left: real part of $t_\pm(s)$ as a function of $s$. It is seen that $t_\pm(s)$ acquires a finite imaginary part for $s \in (s_-,s_+)$, $s_\pm = (M \pm -m)^2$. Right: Movement of $t_\pm(s)$ on the complex plane for varying $s$. In particular, it shown how $t_-(s)$ slides from the upper to the lower part of the complex plane through the left of the branch point $4m^2$, with crossing the cut extending from this point to infinity.\label{fig:tminmax}}
\end{figure}

\begin{figure}\centering
\input{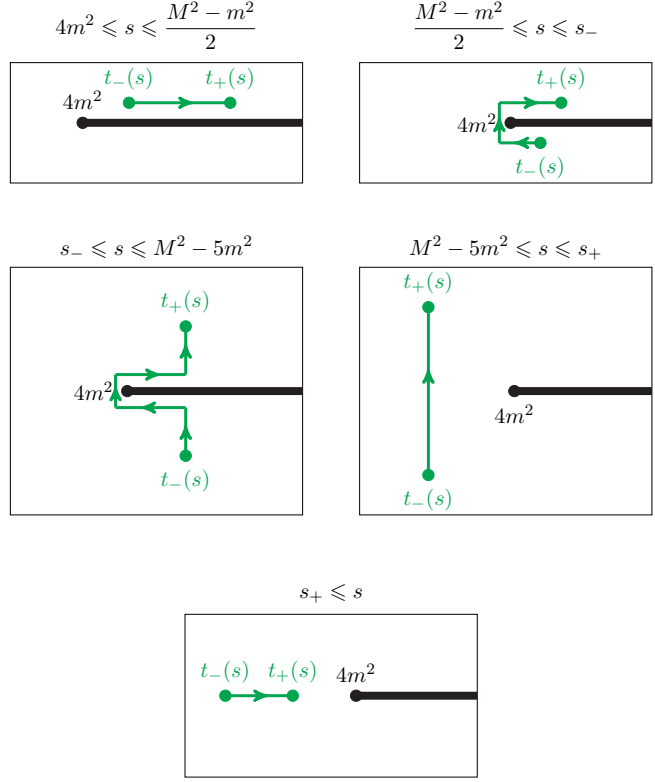}
\caption{Location of the endpoints $t_\pm(s)$ for different values of $s$ with respect to the branch points $4m^2$ and the cut. Also shown is the path deformation discussed in the text.\label{fig:endpoints}}
\end{figure}

The most subtle point in the solution of the integral equations is related to the function $\widetilde{a}(s)$, Eq.~\eqref{eq:sols:ft}, and the endpoints of its defining integral, $t_\pm(s)$. The real part of $t_\pm(s)$ is shown in Fig.~\ref{fig:tminmax} as a function of $s$. We are considering here the physical decay $X \to 3\pi$. For values of the decaying particle mass such that $m_X < 3m$, the cuts of $\widetilde{a}(s)$ functions lie in a region $\text{Re}\, s < 4m^2$, and thus it does not overlap with the RHC, $s > 4m^2$. For physical values $m_X \geqslant 3m$, the cuts would seem to overlap (see Fig.~\ref{fig:tminmax}), but the prescription $m_X^2 + i \epsilon$ \cite{Mandelstam:1960zz} allows the complex cuts of $\tilde{f}(s)$ to be unambiguously separated from the RHC, as shown in the right panel of Fig.~\ref{fig:tminmax}. In practice, this analytic continuation in $m_X^2$ implies that the $t$ integration contour in Eq.~\eqref{eq:sols:ft} must be deformed such that it does not cross the $s$ integration contour, {\it i.e.}, the right-hand cut, $4m^2 \leqslant s < \infty$. A possible such deformation is shown in Fig.~\ref{fig:endpoints}. Thus, to compute $\widetilde{a}(s)$ from Eq.~\eqref{eq:sols:ft}, one possibility is to evaluate the integrand in the $t$ complex plane, since the kernel $K(s,t)$ is analytically known, and either of Eqs.~\eqref{eq:sols:f_no} and \eqref{eq:sols:f_o} allow to compute the function $\hat{a}(t)$ in the complex plane.

We discuss now another possibility, which allows to analytically compute the integrals in $t$. This is achieved by inserting the representation for $\hat{a}(s)$ given in Eq.~\eqref{eq:sols:f_no} into the definition of $\widetilde{a}(s)$, Eq.~\eqref{eq:sols:ft}. By doing so, and reversing the integration order, we get:
\begin{equation}\label{eq:sols:ft_ins}
\widetilde{a}(s) = Q(s) + \int_{s_\text{th}}^{\infty} \mathrm{d}s' \; \text{Disc}\, a(s') \; \mathcal{L}(s,s')~,
\end{equation}
where $Q(s)$ is a polynomial,
\begin{equation}
Q(s) = \int_{t_-(s)}^{t_+(s)} \mathrm{d}t\, K(s,t) \; \widetilde{P}(t)~,
\end{equation}
and $\mathcal{L}(s,s')$ is given by:
\begin{equation}
\mathcal{L}(s,s') = \frac{1}{\pi} \int_{t_-(s)}^{t_+(s)} \mathrm{d}t\, \frac{t^m K(s,t)}{{s^{\prime}}^m (s^\prime - t)} \equiv \overline{\mathcal{L}}(s,s') + K(s,s') \; L(s,s')~,
\end{equation}
where the function $L(s,s')$ is given by:
\begin{equation}
L(s,s') = \frac{1}{\pi} \int_{t_-(s)}^{t_+(s)} \mathrm{d}t \frac{1}{s' - t} =
\log\left( s' - t_-(s) \right) -
\log\left( s' - t_+(s) \right)~.
\end{equation}
The advantage of computing $\widetilde{a}(s)$ through Eq.~\eqref{eq:sols:ft_ins} instead of Eq.~\eqref{eq:sols:ft} is two-fold. First, in this way the function $\hat{a}(s)$ [through its discontinuity, $\text{Disc}\, \hat{a}(s)$] is needed only along the RHC. Second, and more importantly, what is achieved with this decomposition is the isolation of the effect of the necessary contour deformation into a single function, $L(s,s')$. We can explicitly separate the function $L(s,s')$ as follows:
\begin{equation}
L(s,s') \equiv \overline{L}(s,s') + 2i \; \theta(s-s_A) \, \theta(s_B - s) \, \theta\left( \text{Re}[t_-(s)] - s' \right)~,
\end{equation}
and it can be seen that the second term is the one that arises from the deformation of the contour. This piece is the one that gives rise to a well-known \cite{Kambor:1995yc,Gasser:2018qtg} square-root type singularity\footnote{Generally speaking, the power of the singularity, $\left( \sqrt{s-s_-} \right)^\ell$ for some integer $\ell \geqslant 1$, increases with the isobar total angular momentum. The minimum value one find is $\ell = 1$.} of the $\widetilde{a}(s)$ function for $s \to (M-m)^2$.\footnote{We remind here that the functions that we have defined as free of kinematical singularities are actually the $\hat{a}(s)$ ones. One could still say that the $\widetilde{a}(s)$ enter into the discontinuity of $a(s)$, thus making it singular. This argument, however, is not correct, since the discontinuity is $\propto \hat{a}(s+i\epsilon) - \hat{a}(s-i\epsilon)$. The term that is actually diverging is $\hat{a}(s-i\epsilon)$, and the physical amplitude, $\hat{a}(s+i\epsilon)$, remains finite.} The integration limits for the $t'$ integral are $t_{\pm}(s)$, so, if no deformation needs to be done, the interval length is $t_+(s) - t_-(s)  \propto \sqrt{s-s_-}$, which would, in principle, cancel the singularity. By deforming the contour one is including an integral which has not this length, and thus a singularity arises. Care must be taken in dealing with this singularity when a numerical solution of the integral equation system is attempted.

Let us consider how to numerically cope with this singularity, for example, in the integral $I_n(s)$, Eq.~\eqref{eq:sols:f_o_disp}, necessary for the calculation of $\hat{a}(s)$ by means of Eq.~\eqref{eq:sols:f_o}. For compactness, we define the function $\beta(s)$:
\begin{equation}
    s^{-n} I_n(s) =
    \frac{1}{\pi} \int_{4m^2}^{\infty} \mathrm{d} s' \frac{\sin\delta(s') \, \widetilde{a}(s')}{\left\lvert \Omega(s') \right\rvert {s'}^n (s'-s)}
    \equiv
                  \int_{4m^2}^{\infty} \mathrm{d} s' \frac{\widetilde{b}(s')}{s'-s}~.
\end{equation}
Let us consider, as the simplest case, a singularity of $\widetilde{a}(s)$ such as:
\begin{equation}
    \widetilde{a}(s) = \frac{\alpha}{\sqrt{s_- - s'}} + \cdots~,
\end{equation}
which in turn allows one to deduce:
\begin{equation}
    \widetilde{b}(s) = \frac{\beta}{\sqrt{s_- - s'}} + \cdots~.
\end{equation}
The explicit value of $\alpha$ can be computed (in each iteration along the solution procedure) from the integral representations of $\widetilde{a}(s)$, and $\beta$ can be computed from the knowledge of $B_0$ and the functions $\Omega(s)$ and $\delta(s)$. Then, one can write:
\begin{align}
    s^{-n}I_n(s) & = \int_{4m^2}^{\infty} \mathrm{d} s' \frac{\widetilde{b}(s')}{s'-s} =
    \pvint_{4m^2}^{\infty} \mathrm{d} s' \frac{1}{s'-s} \left( \beta(s') - \frac{\beta_0}{\sqrt{s_- - s'}} \right) \\
    & + i \pi \left(  \beta(s) - \frac{\beta_0}{\sqrt{s_- - s}} \right) + \beta_0\, \beta_1(s)~. \nonumber
\end{align}
Above, $\pvint$ stands for the principal value integral. The first integral is suitable for numerical integration, and the function $\beta_1(s)$, the reminder integral, can be explicitly computed, and it is indeed analytical at $s=s_-$. Furthermore, the singularity of $\beta(s)$ cancels with the term $\beta_0/\sqrt{s_- - s}$ inside the parenthesis. Similar methods can be applied to deal with the singularities of $\tilde{a}(s)$ in the different integrals in which it appears.

\bibliographystyle{apsrev4-1}
\bibliography{refs}

\end{document}